\documentclass[acmsmall,screen]{acmart}

\startPage{1}

\setcopyright{none}

\usepackage{booktabs}  
\usepackage{subcaption} 

\usepackage{amsmath,amsthm,stmaryrd,mathtools}
\usepackage{listings}
\usepackage{mathpartir}
\usepackage{proof}
\usepackage{subcaption}
\usepackage{hyperref}
\usepackage{breakurl}
\usepackage{xspace}
\usepackage{multicol}
\usepackage{multirow}
\usepackage{enumitem}
\newcommand{\toolname}{{\sc zeus}\xspace}
\usepackage{makecell}
\usepackage{microtype}

\begin{document}

\title{Program Equivalence for Assisted Grading of Functional Programs (Extended Version)}         

\author{Joshua Clune}
                                        
\affiliation{
  \institution{Carnegie Mellon University}   
  \country{United States of America}
}
\email{josh.seth.clune@gmail.com}      

\author{Vijay Ramamurthy}
                                        
\affiliation{
  \institution{Carnegie Mellon University}      
  \country{United States of America}
}
\email{vrama628@gmail.com}          

\author{Ruben Martins}
\affiliation{
  \institution{Carnegie Mellon University}    
  \country{United States of America}
}
\email{rubenm@andrew.cmu.edu}         

\author{Umut A. Acar}
                                        
\affiliation{
  \institution{Carnegie Mellon University} 
  \country{United States of America}
}
\email{umut@cs.cmu.edu}         

\begin{abstract}

In courses that involve programming assignments, giving meaningful feedback to students is an important challenge. Human beings can give useful feedback by manually grading the programs but this is a time-consuming, labor intensive, and usually boring process.   Automatic graders can be fast and scale well but they usually provide poor feedback. Although there has been research on improving automatic graders, research on scaling and improving human grading is limited. 

We propose to scale human grading by augmenting the manual grading process with an equivalence algorithm that can identify the equivalences between student submissions.  This enables human graders to give targeted feedback for multiple student submissions at once. Our technique is conservative in two aspects. First, it identifies equivalence between submissions that are algorithmically similar, e.g., it cannot identify the equivalence between quicksort and mergesort. Second, it uses formal methods instead of clustering algorithms from the machine learning literature. This allows us to prove a soundness result that guarantees that submissions will never be clustered together in error. Despite only reporting equivalence when there is algorithmic similarity and the ability to formally prove equivalence, we show that our technique can significantly reduce grading time for thousands of programming submissions from an introductory functional programming course.

\end{abstract}



\maketitle

\section{Introduction}

There have been many efforts to develop techniques for automated reasoning of programming assignments at scale. This has lead to the rise of automatic graders, programs that take in a set of student submissions and output grades or feedback for those submissions without requiring any human input. While recent years have yielded substantial improvements in automatic grading techniques \cite{gulwani-clustering-pldi18,wang-sarfgen-pldi18,kaleeswaran-coderassist-fse16,singh-autograder-pldi13,liu-grading-formal-icse19,perry-semcluster-pldi19}, automatic graders are still more limited in the feedback they can provide than human graders. 

This creates a trade-off between scale and quality. For small courses, it makes sense to utilize human graders in order to provide the best feedback possible. For Massive Open Online Courses, human involvement in grading all submissions is often logistically impossible, so it makes sense to use automatic graders. But neither option is ideal for large, in-person, introductory functional courses. When introductory functional courses use automatic graders, it hurts the students because they receive less targeted feedback, and it can hurt the teaching staff to lose a valuable avenue for addressing uncommon misunderstandings. But when introductory functional courses use human graders, it creates a large burden on the teaching staff, and it may require capping the size of the class, hurting students by limiting their opportunity to take the class.

To provide an option that eases the cost of human grading without sacrificing feedback quality, we propose a method of enabling human graders to give targeted feedback to multiple students at once. Our approach takes a pair of expressions submitted by students and deconstructs them simultaneously to build up a formula that is valid only if the expressions are equivalent. This pairwise equivalence test is used to cluster student submissions into buckets for which all submissions can be graded and given feedback simultaneously. Our approach recognizes expressions as equivalent by finding equivalences in each expression's subexpressions. To do this, it uses a variety of inference rules to simultaneously deconstruct the expressions down to their atomic subexpressions. It then outputs formulas that are valid only if the atomic subexpressions are equivalent. Finally, our inference rules recursively use the formulas of these subexpressions as subformulas to build up a larger formula that indicates the equivalence of the overall expression. This final formula's validity can be checked by an SMT Solver to determine whether the two expressions are equivalent.

A central benefit of our approach is that when two expressions are recognized as equivalent, this fact does not merely reflect that the two expressions produce the same outputs on shared inputs. In input/output grading, the correctness of code is determined entirely by whether a student submission produces correct outputs when given a large and diverse set of inputs. But in our approach, all equivalences arise from similarities in subexpressions, so equivalences found by our technique are discoverable only due to underlying algorithmic similarities. This enables instructors to give feedback based not only on whether a problem was solved correctly, but based on the algorithmic decisions that were involved in the student's solution.

Three primary factors that impact the grading and feedback of student programs are correctness, algorithmic approach, and style. While our approach is meant to enable providing better feedback concerning algorithmic approach, as opposed to simply providing feedback concerning correctness as in input/output grading, evaluating style is outside of the scope of our technique. For that reason, we believe that our approach is best utilized in conjunction with the methods courses already use to evaluate style. For courses already doing automatic grading, this should not be an issue because if they are already doing automatic grading, they are already automatically doing style checking, and can, therefore, use that in conjunction with our approach to provide all of the same style feedback the course already provided, but additionally provide human feedback for algorithmic content.

For courses already doing fully human grading, even if it is still necessary to grade each assignment individually to address style concerns, we believe our approach can make it possible to better allocate human resources for the grading process. A grader focusing entirely on one or two large buckets can be more efficient by not being forced to figure out which common approach is being taken by every individual submission. This can help the grader more quickly move on from understanding the student’s solution to addressing any style concerns, and it also helps ensure fairer grading in guaranteeing that the same grader will grade all similar submissions. A grader focusing entirely on grading submissions that were clustered with few if any other programs can anticipate ahead of time that their grading will likely require providing more frequent and/or detailed comments. This can enable course staffs to give more submissions to graders of large buckets, easing the burden of singleton/small bucket graders.

The differences between our approach and other state-of-the-art automatic graders and clustering techniques~\cite{gulwani-clustering-pldi18,wang-sarfgen-pldi18,perry-semcluster-pldi19} stem from differences in motivation. Since each bucket generated by our approach is meant to be graded by a human, it is more important for our technique to distinguish nonequivalent submissions than to ensure that all equivalent submissions are placed in the same bucket. Ensuring that all equivalent submissions are placed in the same bucket reduces time spent grading equivalent programs, enabling instructors to spend more time giving detailed feedback. This is an important goal, but it is of lower priority than preserving the accuracy of human feedback because it does not matter how detailed feedback is if it does not apply to the student to whom it is given. To secure the accuracy of human feedback while using our approach, we guarantee the correctness of our technique’s recognized equivalences by proving a soundness theorem that states that if our technique recognizes two expressions as equivalent, they necessarily exhibit identical behavior.

In summary, the contributions of our paper are as follows:
\setitemize{noitemsep,topsep=0pt,parsep=0pt,partopsep=0pt}
\begin{itemize}
    \item We define an effective and efficient technique for identifying equivalences between purely functional programs. The technique's design ensures that only algorithmically similar programs will be recognized as equivalent.
    \item We prove the soundness of this technique, showing that if our approach identifies an equivalence between two expressions, then the two expressions must exhibit identical behavior.
    \item We implement our approach in a tool called \toolname and demonstrate its effectiveness in assisting the grading of more than 4,000 student submissions from a functional programming course taught at the college level in Standard ML.
\end{itemize}

\section{Motivating Examples}

Our approach is meant to cluster expressions that are algorithmically similar, but potentially syntactically different. In this section, we show two examples of similar implementations of the same function that are successfully identified by our tool as equivalent, and describe one example in which two solutions to a task are not recognized as equivalent due to algorithmic dissimilarities.

\begin{figure}[h] \centering
\begin{minipage}[t]{0.45\textwidth}
\vspace{0pt}
\centering
\begin{tabular}{c}
\begin{lstlisting}
fun add_opt x y =
  case (x, y) of
    (SOME m, SOME n) =>
      SOME (m + n)
  | (NONE, _) => NONE
  | (_, NONE) => NONE
\end{lstlisting}
\end{tabular}
\end{minipage}
\begin{minipage}[t]{0.45\textwidth}
\vspace{0pt}
\centering
\begin{tabular}{c}
\begin{lstlisting}
fun bind a f =
  case a of
    SOME b => f b
  | NONE => NONE

val return = SOME

fun add_opt x y =
  bind x (fn m =>
  bind y (fn n =>
    return (m + n)
  ))
\end{lstlisting}
\end{tabular}
\end{minipage}
\caption{Two implementations of adding two optional numbers}
\label{fig:motivating example add_opt}
\end{figure}

Figure \ref{fig:motivating example add_opt} contains two functions that take in two \texttt{int options} as input, and adds the \texttt{ints} in the \texttt{options} if possible, returning \texttt{NONE} otherwise. The right expression's conditional logic is modeled after Haskell-style monads, interacting with the higher order bind function to case on \texttt{x} first, and then potentially \texttt{y} depending on the value of \texttt{x}, whereas the left expression cases on \texttt{x} and \texttt{y} simultaneously. Still, our approach is able to fully encode both expressions' conditional logic structures and produce a valid formula. A demonstration of how our approach specifically encodes these conditional logic structures is included in Section \ref{section:operation}.

\begin{figure}[h] \centering
\begin{minipage}[t]{0.45\textwidth}
\vspace{0pt}
\centering
\begin{tabular}{c}
\begin{lstlisting}
fun split [] = ([], [])
  | split [x] = ([x], [])
  | split (x::y::L) = 
      let
        val (A, B) = split L
      in
        (x::A, y::B)
      end

fun merge([], L) = L
  | merge(L, []) = L
  | merge(x::xs, y::ys) =
      if x < y
      then x :: merge (xs, y::ys)
      else y :: merge (x::xs, ys)

fun msort [] = []
  | msort [x] = [x]
  | msort L =
      let
        val (A, B) = split L
      in
        merge(msort A, msort B)
      end
\end{lstlisting}
\end{tabular}
\end{minipage}
\begin{minipage}[t]{0.45\textwidth}
\vspace{0pt}
\centering
\begin{tabular}{c}
\begin{lstlisting}
fun split [] = ([], [])
  | split (x::xs) =
      case xs of
          [] => ([x], [])
        | (y::ys) =>
            let
              val (A, B) = split ys
            in
              (x::A, y::B)
            end

fun merge (l1, l2) =
  case l1 of
    [] => l2
  | x::xs =>
    case l2 of
      [] => l1
    | y::ys =>
      if x < y
      then x :: merge (xs, l2)
      else y :: merge (l1, ys)

fun msort [] = []
  | msort [x] = [x]
  | msort L =
      let
        val (A, B) = split L
      in
        merge(msort A, msort B)
      end
\end{lstlisting}
\end{tabular}
\end{minipage}
\caption{Two implementations of mergesort}
\label{fig:motivating example msort}
\end{figure}

Figure \ref{fig:motivating example msort} contains two functions that implement mergesort. The left implementation uses a style that emphasizes pattern matching on input arguments while the right implementation uses a style that emphasizes nesting binding structures. Despite their syntactic differences, both functions implement the same underlying algorithm. Therefore, our approach recognizes them as equivalent.

Our approach is not intended to cluster programs just by correctness, or final input/output behavior, but by structure. This enables our approach to distinguish between correct submissions that use different algorithms. For instance, one of the benchmarks we use in Section \ref{section:results} to evaluate our tool is a task called slowDoop. The goal of this task is to take in an arbitrary list $L$ and return a list in which all elements in $L$ appear exactly once. Consider a similar task in which the goal is the same but has the added stipulation that the final list must be sorted. A reasonable O($n^2$) solution to this task would be to iterate over $L$, only keeping elements that do not appear later in the list, and then sort the result. But a better O($n$log$n$) solution would be to first sort $L$, and then iterate over the resulting list once to remove duplicate elements. While correct implementations of these algorithms are identical from an input/output perspective, our approach would cluster them separately, and we believe that they merit different feedback.

\section{LambdaPix}

Our approach operates over a language which we call LambdaPix. LambdaPix is designed to be a target for transpilation from functional programming languages such as Standard ML, OCaml, or Haskell. Our techniques apply to purely functional programs only and do not allow for state (e.g., references) but are otherwise unrestricted and make no further assumptions about the programs.
In this section, we present the syntax and semantics for LambdaPix.

\begin{figure}[h]
\begin{align*}
\small{
\begin{array}{l c c l@{\qquad} l}
base\ types & b & ::= & int\ |\ boolean\\
types & \tau  & ::= & b & base\ type \\
    &       & |    & \delta & data\ type \\
    &       & |    & \{\ell_1 : \tau_1, \ldots, \ell_n : \tau_n \} & product\ type \\
    &       & |   & \tau_1 \rightarrow \tau_2 & function\ type \\
injection\ labels & i & ::= & \textnormal{label}_1\ |\ \textnormal{label}_2\ |\ \ldots\\
patterns & p & ::= & \_ & wildcard\ pattern \\
    &   &  |   & x & variable\ pattern \\
    &   &  |   & \{\ell_1 = p_1, \ldots, \ell_n = p_n \} & record\ pattern \\
    &   &  |   & x\ \texttt{as}\ p & alias\ pattern \\
    &   &  |   & c & constant\ pattern \\
    &   &  |   & i \cdot p & injection\ pattern\ (with\ argument) \\
    &   &  |   & i & injection\ pattern\ (without\ argument)\\
primitive\ operations & o & ::= & +\ |\ -\ |\ *\ |\ <\ |\ >\ |\ \leq\ |\ \geq\\
expressions & e & ::= & c & constant \\
    &   &  |   & x & variable \\
    &   &  |   & \{\ell_1 = e_1, \ldots, \ell_n = e_n \} & record \\
    &   &  |   & e \cdot \ell_i & projection \\
    &   &  |   & i \cdot e & injection\ (with\ argument) \\
    &   &  |   & i & injection\ (without\ argument) \\
    &   &  |   & \texttt{case}\ e\ \{ p_1 . e_1 \mid \ldots \mid p_n . e_n \}  & case\ analysis \\
    &   &  |   & \lambda x . e  & abstraction \\
    &   &  |   & e_1\ e_2 & application \\
    &   &  |   & \texttt{fix}\ x\ \texttt{is}\ e & fixed\ point \\
    &   &  |   & o & primitive\ operation
\end{array}
}
\end{align*}
\caption{The syntax of LambdaPix}
\label{fig:syntax}
\end{figure}

We give the syntax for LambdaPix in Figure \ref{fig:syntax}.
Arbitrary labeled product types are supported as labeled records.
For sum types and recursive types, LambdaPix is defined over an arbitrary
fixed set of algebraic data types, with associated injection labels. We use meta-variables $x$, $y$, and $z$ (and variants) to range over an unspecified set of variables.

\subsection{Static Semantics}

We assume an arbitrary fixed set of disjoint algebraic data types with unique associated injection labels (by unique, it is meant that there are no shared injection labels between distinct data types).
In particular, we assume a fixed set of judgments of the form
$i : \tau \hookrightarrow \delta$ for injection labels that take in an argument of type $\tau$ to produce an expression of data type $\delta$, and a fixed set of judgments of the form $i : \delta$ for injection labels of data type $\delta$ that do not take in an argument.
We take $i : \tau \hookrightarrow \delta$ to mean that
the type $\delta$ has a label $i$ which accepts an argument of
type $\tau$, and we take $i : \delta$ to mean that the type $\delta$ has a label $i$ that does not accept an argument.
Note that by allowing $\tau$ to contain instances of $\delta$,
this data type system affords LamdbaPix a form of inductive types.

\newcounter{pattyrule}
\newcommand{\definepattyrule}[1]{\refstepcounter{pattyrule}\label{patty:#1}\textsc{PatTy}_{\thepattyrule}}
\newcommand{\pattyrule}[1]{\ensuremath{\textsc{PatTy}_\text{\ref{patty:#1}}}}
\begin{figure}[h]
{\small
\centering
\begin{mathpar}
\infer[\definepattyrule{wildcard}]{\_ :: \tau \dashv}{}

\infer[\definepattyrule{variable}]{x :: \tau \dashv x : \tau}{}

\infer[\definepattyrule{record}]{
  \{l_1 = p_1, \ldots, l_n = p_n \} ::
  \{\ell_1 : \tau_1, \ldots, \ell_n : \tau_n \}
  \dashv \Gamma_1 \ldots \Gamma_n
}{
  p_1 :: \tau_1 \dashv \Gamma_1 &
  \ldots &
  p_n :: \tau_n \dashv \Gamma_n
}

\infer[\definepattyrule{alias}]{x\ \texttt{as}\ p :: \tau \dashv \Gamma, x : \tau}{
  p :: \tau \dashv \Gamma
}

\infer[\definepattyrule{constant}]{c :: b \dashv}{}

\infer[\definepattyrule{injection (without arg)}]{i :: \delta \dashv}{i : \delta}

\infer[\definepattyrule{injection (with arg)}]{i \cdot p :: \delta \dashv \Gamma}{
  i : \tau \hookrightarrow \delta &
  p :: \tau \dashv \Gamma
}
\end{mathpar}
}
\caption{Pattern typing in LambdaPix}
\label{fig:pattern_typing}
\end{figure}

Figure \ref{fig:pattern_typing} defines an auxiliary judgment used in the
typechecking of case expressions.
This pattern typing judgment $p :: \tau \dashv \Gamma$ defines that expressions
of type $\tau$ can be matched against the pattern $p$, and that doing so
produces new variable bindings whose types are captured in $\Gamma$.

\newcounter{tyrule}
\newcommand{\definetyrule}[1]{\refstepcounter{tyrule}\label{ty:#1}\textsc{Ty}_{\thetyrule}}
\newcommand{\tyrule}[1]{\ensuremath{\textsc{Ty}_\text{\ref{ty:#1}}}}
\begin{figure}[h]
{\small
\centering
\begin{mathpar}
\infer[\definetyrule{constant}]{\Gamma \vdash c : b}{}

\infer[\definetyrule{variable}]{\Gamma, x : \tau \vdash x : \tau}{}

\infer[\definetyrule{record}]{\Gamma \vdash
  \{\ell_1 = e_1, \ldots, \ell_n = e_n \} :
  \{\ell_1 : \tau_1, \ldots, \ell_n : \tau_n \}
}{
  \Gamma \vdash e_1 : \tau_1 &
  \ldots &
  \Gamma \vdash e_n : \tau_n
}

\infer[\definetyrule{projection}]{
  \Gamma \vdash e \cdot \ell_i : \tau_i
}{
  \Gamma \vdash e : \{\ldots, \ell_i : \tau_i, \ldots \}
}

\infer[\definetyrule{injection (without arg)}]{\Gamma \vdash i : \delta}{i : \delta}

\infer[\definetyrule{injection (with arg)}]{\Gamma \vdash i \cdot e : \delta}{
  i : \tau \hookrightarrow \delta &
  \Gamma \vdash e : \tau
}

\infer[\definetyrule{case}]{\Gamma \vdash
  \texttt{case}\ e\ \{ p_1 . e_1 \mid \ldots \mid p_n . e_n \} : \tau'
}{
  \Gamma \vdash e : \tau &
  p_1 :: \tau \dashv \Gamma_1 &
  \Gamma, \Gamma_1 \vdash e_1 : \tau' &
  \ldots &
  p_n :: \tau \dashv \Gamma_n &
  \Gamma, \Gamma_n \vdash e_n : \tau'
}

\infer[\definetyrule{lambda}]{
  \Gamma \vdash \lambda x . e : \tau_1 \rightarrow \tau_2
}{
  \Gamma, x : \tau_1 \vdash e : \tau_2
}

\infer[\definetyrule{application}]{
  \Gamma \vdash e_1\ e_2 : \tau_2
}{
  \Gamma \vdash e_1 : \tau_1 \rightarrow \tau_2 &
  \Gamma \vdash e_2 : \tau_1
}

\infer[\definetyrule{fix}]{
  \Gamma \vdash \texttt{fix}\ x\ \texttt{is}\ e : \tau
}{
  \Gamma, x : \tau \vdash e : \tau
}

\infer[\definetyrule{built-in function}]{
  \Gamma, o : \tau_1 \rightarrow \tau_2 \vdash o : \tau_1 \rightarrow \tau_2
}
{}
\end{mathpar}
}
\caption{Expression typing in LambdaPix}
\label{fig:expression_typing}
\end{figure}

Figure \ref{fig:expression_typing} defines typing for expressions in LambdaPix.
\begin{definition}[Well-formed]
A LambdaPix expression $e$ is well-formed if there exists a type $\tau$ such
that $\Gamma_{\textnormal{initial}} \vdash e : \tau$, where $\Gamma_{\textnormal{initial}}$ only contains the typing judgments for primitive operations.
\end{definition}

Not captured in the type system of LambdaPix are the following two
restrictions:
\begin{itemize}
\item No variable may appear more than once in a pattern.
\item The patterns of a case expression must be exhaustive.
\end{itemize}

\subsection{Dynamic Semantics}
\newcommand{\matches}[3]{#1 \sslash #2 \dashv #3}
\newcommand{\nomatch}[2]{#1 \mathbin{\setminus \mkern-10mu \sslash} #2}
\newcommand{\stepsto}[2]{#1 \mapsto #2}
\newcommand{\val}[1]{#1\ \mathsf{val}}

Here we define how LambdaPix expressions evaluate.
We define evaluation as a small-step dynamic semantics where the judgment
$\stepsto{e}{e'}$ means that $e$ steps to $e'$ and the judgment $\val{e}$
means that $e$ is a value and doesn't step any further.
LambdaPix enjoys progress and preservation.
\begin{definition}[Progress and Preservation]
For any typing context $\Gamma$ and expression $e$
such that $\Gamma \vdash e : \tau$ it is either the case that $\val{e}$ or there exists an $e'$ such that $\Gamma \vdash e' : \tau$ and $\stepsto{e}{e'}$.
\end{definition}

\noindent LambdaPix also enjoys the finality of values: it is never the case
that both $\stepsto{e}{e'}$ and $\val{e}$.

\newcounter{matchrule}
\newcommand{\definematchrule}[1]{\refstepcounter{matchrule}\label{match:#1}\textsc{Match}_{\thematchrule}}
\newcommand{\matchrule}[1]{\ensuremath{\textsc{Match}_\text{\ref{match:#1}}}}
\begin{figure}[h]
{\small
\centering
\begin{mathpar}
\infer[\definematchrule{wildcard}]{\matches{v}{\_}{}}{}

\infer[\definematchrule{variable}]{\matches{v}{x}{v/x}}{}

\infer[\definematchrule{constant1}]{
  \matches{c_1}{c_2}{}
}{
  c_1 = c_2
}

\infer[\definematchrule{constant2}]{
  \nomatch{c_1}{c_2}
}{
  c_1 \not= c_2
}

\infer[\definematchrule{record1}]{
  \matches{
    \{\ell_1=v_1, \ldots, \ell_n=v_n\}
  }{
    \{\ell_1=p_1, \ldots, \ell_n=p_n\}
  }{
    B_1 \ldots B_n
  }
}{
  \matches{v_1}{p_1}{B_1} &
  \ldots &
  \matches{v_n}{p_n}{B_n}
}

\infer[\definematchrule{record2}]{
  \nomatch{
    \{\ell_1=v_1, \ldots, \ell_n=v_n\}
  }{
    \{\ell_1=p_1, \ldots, \ell_n=p_n\}
  }
}{
  \nomatch{v_i}{p_i}
}

\infer[\definematchrule{alias1}]{
  \matches{v}{x\ \texttt{as}\ p}{B, v/x}
}{
  \matches{v}{p}{B}
}

\infer[\definematchrule{alias2}]{
  \nomatch{v}{x\ \texttt{as}\ p}
}{
  \nomatch{v}{p}
}

\infer[\definematchrule{injection (without args)1}]
{\matches{i}{i}{}}{}

\infer[\definematchrule{injection (without args)2}]
{\nomatch{i_1}{i_2}{}}{i_1 \not= i_2}

\infer[\definematchrule{injection (with args)1}]{
  \matches{i \cdot v}{i \cdot p}{B}
}{
  \matches{v}{p}{B}
}

\infer[\definematchrule{injection (with args)2}]{
  \nomatch{i_1 \cdot v}{i_2 \cdot p}
}{
  i_1 \not= i_2
}

\infer[\definematchrule{injection (with args)3}]{
  \nomatch{i \cdot v}{i \cdot p}
}{
  \nomatch{v}{p}
}

\infer[\definematchrule{injection with/without args}]{
  \nomatch{i_1 \cdot v}{i_2}
}{}

\infer[\definematchrule{injection without/with args}]{
  \nomatch{i_1}{i_2 \cdot p}
}{}
\end{mathpar}
}
\caption{Pattern matching in LambdaPix}
\label{fig:matching}
\end{figure}
To define evaluation
we first define two helper judgments to deal with pattern matching (Figure
\ref{fig:matching}).
The judgment $\matches{v}{p}{B}$ means the value $v$ matches to
the pattern $p$ producing $B$, where $B$ is a set of bindings of the form $v' / x$ that indicate the value $v'$ is bound to the variable $x$. The judgment $\nomatch{v}{p}$ means the expression $v$ does not match to the pattern $p$.
It is assumed as a precondition to these judgements that $\val{v}$, $\vdash v : \tau$, and $p :: \tau$.
Pattern matching in LambdaPix enjoys the property that for any $v$ and $p$ satisfying the above preconditions it is either the case that there exist bindings $B$ such that $\matches{v}{p}{B}$, or $\nomatch{v}{p}$.
It is never simultaneously the case that $\matches{v}{p}{B}$ and
$\nomatch{v}{p}$.
\newcounter{dynrule}
\newcommand{\definedynrule}[1]{\refstepcounter{dynrule}\label{dyn:#1}\textsc{Dyn}_{\thedynrule}}
\newcommand{\dynrule}[1]{\ensuremath{\textsc{Dyn}_\text{\ref{dyn:#1}}}}
\newcounter{bigdynrule}
\newcommand{\definebigdynrule}[1]{\refstepcounter{bigdynrule}\label{bigdyn:#1}\textsc{BigDyn}_{\thebigdynrule}}
\newcommand{\bigdynrule}[1]{\ensuremath{\textsc{BigDyn}_\text{\ref{bigdyn:#1}}}}
\newcommand{\evaluatesto}[2]{#1 \Mapsto #2}
\begin{figure}[h]
{\small
\centering
\begin{mathpar}
\infer[\definedynrule{constant}]{\val{c}}{}

\infer[\definedynrule{record1}]{
    \stepsto{
        \{\ldots, \ell_i = e_i, \ldots\}
    }{
        \{\ldots, \ell_i = e'_i, \ldots\}
    }
}{
    \val{e_1} & \val{e_2} & \ldots & \val{e_{i-1}} & 
    \stepsto{e_i}{e'_i} &
}

\infer[\definedynrule{record2}]{
    \val{\{\ell_1 = e_1, \ldots, \ell_n = e_n \}}
}{
    \val{e_1} & \ldots & \val{e_n}
}

\infer[\definedynrule{projection1}]{
  \stepsto{e \cdot \ell_i}{e' \cdot \ell_i}
}{
  \stepsto{e}{e'}
}

\infer[\definedynrule{projection2}]{
  \stepsto{\{\ldots, \ell_i = e_i, \ldots \} \cdot \ell_i}{e_i}
}{
  \val{\{\ldots, \ell_i = e_i, \ldots \}}
}

\infer[\definedynrule{injection (with args)1}]{
  \stepsto{i \cdot e}{i \cdot e'}
}{
  \stepsto{e}{e'}
}

\infer[\definedynrule{injection (with args)2}]{
  \val{i \cdot e}
}{
  \val{e}
}

\infer[\definedynrule{injection (without args)}]
{\val{i}}{}

\infer[\definedynrule{case1}]{
  \stepsto{
    \texttt{case}\ e\ \{ p_1 . e_1 \mid \ldots \mid p_n . e_n \}
  }{
    \texttt{case}\ e'\ \{ p_1 . e_1 \mid \ldots \mid p_n . e_n \}
  }
}{
  \stepsto{e}{e'}
}

\infer[\definedynrule{case2}]{
  \stepsto{
    \texttt{case}\ e\ \{\ldots \mid p_i . e_i \mid \ldots \}
  }{
    [B]e_i
  }
}{
  \val{e} & 
  \nomatch{e}{p_1} &
  \ldots &
  \nomatch{e}{p_{i-1}} &
  \matches{e}{p_i}{B}
}

\infer[\definedynrule{lambda}]{\val{\lambda x . e}}{}

\infer[\definedynrule{application1}]{
  \stepsto{e_1\ e_2}{e'_1\ e_2}
}{
  \stepsto{e_1}{e'_1}
}

\infer[\definedynrule{application2}]{
  \stepsto{e_1\ e_2}{e_1\ e'_2}
}{
  \val{e_1} &
  \stepsto{e_2}{e'_2}
}

\infer[\definedynrule{beta}]{
  \stepsto{(\lambda x . e)\ e_2}{[e_2/x]e}
}{
  \val{e_2}
}

\infer[\definedynrule{fix}]{
  \stepsto{
    \texttt{fix}\ x\ \texttt{is}\ e
  }{
    [\texttt{fix}\ x\ \texttt{is}\ e/x]e
  }
}{}

\infer[\definedynrule{built-in function}]{
    \val{o}
}{}

\infer[\definedynrule{built-in function2}]{
    \stepsto{o\ e}{e'}
}
{\val {e}}

\infer[\definebigdynrule{val}]{
  \evaluatesto{v}{v}
}{
  \val {v}
}

\infer[\definebigdynrule{step}]{
  \evaluatesto{e}{v}
}{
  \stepsto{e}{e'} &
  \evaluatesto{e'}{v}
}

\end{mathpar}
}
\caption{Dynamic semantics of LambdaPix}
\label{fig:dynamics}
\end{figure}

In Figure \ref{fig:dynamics} we use these helper judgments to define the evaluation judgments. In \dynrule{built-in function2}, $e'$ is meant to be understood as a hard-coded value dependent on the primitive operation $o$. We use these judgments to define what it means for an expression to evaluate
to a value.
We use $\evaluatesto{e}{v}$ to denote that expression $e$ evaluates to value $v$.
In rules \bigdynrule{val} and \bigdynrule{step}, big-step dynamics are defined as the transitive closure of the small-step dynamics.

\section{Sound Equivalence Inferences}

Our approach takes as input two LambdaPix expressions of the same type and outputs a logic formula which is valid only if the two expressions are equivalent. We construct this logic formula by constructing a proof tree of sound equivalence inferences.

\subsection{Logic Formulas}

\newcommand{\forallval}{\overset{\mathsf{val}}{\forall}}

\makeatletter
\newcommand\smtequivnormal[2][]{%
  \ext@arrow 9999{\longleftrightarrowfill@}{#1}{#2}}
\newcommand\longleftrightarrowfill@{%
  \arrowfill@\leftarrow\relbar\rightarrow}
\makeatother

\makeatletter
\newcommand\smtequiv[2][]{%
  \ext@arrow 9999{\longLeftRightarrowfill@}{#1}{#2}}
\newcommand\longLeftRightarrowfill@{%
  \arrowfill@\Leftarrow \Relbar \Rightarrow}
\makeatother

\newcommand{\pathiso}{\leftrightarrow}
\newcommand{\expiso}{\Leftrightarrow}
\newcommand{\whnfstep}{\leadsto}
\newcommand{\whnfbig}{\downarrow}
\newcommand{\fresh}{\ \mathsf{fresh}}

\begin{figure}[h]
\centering
\begin{align*}
\begin{array}{l c c l@{\qquad} l}
\sigma & ::= & t_1 \equiv t_2 & term\ equivalence \\
&     & \sigma_1 \land \sigma_2 & conjunction \\
&     & \sigma_1 \vee \sigma_2 & disjunction\\
&     & \sigma_1 \Rightarrow \sigma_2 & implication \\
&     & \neg{\sigma} & negation
\end{array}
\end{align*}
\caption{Logic Formulas}
\label{fig:propformulas}
\end{figure}

\newcommand{\term}{\ \mathsf{Term}}
\newcounter{termrule}
\newcommand{\definetermrule}[1]{\refstepcounter{termrule}\label{term:#1}\textsc{Term}_{\thetermrule}}
\newcommand{\termrule}[1]{\ensuremath{\textsc{Term}_\text{\ref{term:#1}}}}
\begin{figure}[h]
{\small
\centering
\begin{mathpar}
\infer[\definetermrule{constant}]
{c \term}{}

\infer[\definetermrule{variable}]
{x \term}{}

\infer[\definetermrule{record}]
{\{\ell_1 = t_1, \ldots, \ell_n = t_n \} \term}
{t_1 \term 
&t_2 \term
&\ldots
&t_n \term
}

\infer[\definetermrule{projection}]
{t \cdot \ell_i \term}
{t \term}

\infer[\definetermrule{injection (with arg)}]
{i \cdot t \term}
{t \term}

\infer[\definetermrule{injection (without arg)}]
{i \term}{}

\infer[\definetermrule{wildcard pattern}]
{\_ \term}
{}

\infer[\definetermrule{as pattern}]
{x\ \texttt{as}\ t \term}
{t \term}

\infer[\definetermrule{built-in function}]
{o \term}{}

\infer[\definetermrule{built-in function application}]
{o\ t \term}
{t \term}

\end{mathpar}
}
\caption{Term Judgment}
\label{fig:term}
\end{figure}

Figure \ref{fig:propformulas} defines the form of the formulas generated by our approach.
The leaves of these formulas are equalities between base terms $t$, defined in Figure \ref{fig:term}. These base terms encode three things: LambdaPix values, patterns, and the application of a primitive operation and a value. Encoding all of these things as terms allows a term equivalence to state that either two values are the same, that a value matches with a pattern, or that a primitive operation application yields a value that is equal to another value or matches with a pattern. 

The inclusion of primitive operation applications as base terms is somewhat strange since they are not values in the actual dynamics of LambdaPix, but this inclusion enables the resulting formula to include all of the information pertaining to the theory from which the primitive operation originates. For instance, since the theory of quantifier-free linear integer arithmetic knows that addition is commutative, this inclusion makes it possible for the expressions $\lambda x.\lambda y. (x + y)$ and $\lambda x.\lambda y. (y + x)$ to be recognized as equivalent.

Except when a variable, primitive operation, as pattern, or wildcard pattern is included in one of the terms, term equivalence is identical to syntactic equality. When a primitive operation is included in a term, the specific primitive operation is used to determine how to understand the term equivalence (e.g. $1+2\equiv3$ is a valid term equivalence using the primitive operation "+"). When an \texttt{as} pattern is included in a term equivalence: $x\ \texttt{as}\ e_1 \equiv e_2$, the term equivalence is the same as $x \equiv e_1 \land e_1 \equiv e_2$. When a wildcard pattern is included in a term equivalence: $\_ \equiv e$, the term equivalence can simply be interpreted as "true".

When one or more free variables are included in a formula, they must be resolved to determine the formula's truth. Throughout our approach, contexts are used to keep track of the types of all of a formula's free variables. Expressions can be substituted for variables of the same type in a formula to resolve it (e.g. $[3/x](x\equiv1 \land x \equiv 2)$ yields $3\equiv1 \land 3\equiv2$). A formula is valid if it is true under all possible substitutions of its variables. To denote this, we define a new form of judgment:

\begin{definition}[$\forallval_\Gamma . j$]
If $\Gamma = \vec{x} : \vec{\tau}$, then the judgement $\forallval_\Gamma . j$ holds if for all $\vec{v}$ where $v_i : \tau_i$ and $\val{v_i}$ for all $v_i \in \vec{v}$, it is the case that $[\vec{v}/\vec{x}]j$ holds. Implicitly, although the types of primitive operations are included in $\Gamma_{\textnormal{initial}}$, and therefore $\Gamma$, we omit typings of the form $o : \tau_1 \rightarrow \tau_2$ from $\vec{x} : \vec{\tau}$ so that we do not range over all possible meanings for LambdaPix's primitive operations. Then if $\Gamma$ is a typing context with a mapping for every free variable in a formula $\sigma$, the validity of $\sigma$ is denoted $\forallval_\Gamma . \sigma$.
\label{def:valid}
\end{definition}

The validity of formulas will be what determines whether our approach recognizes two LambdaPix expressions as equivalent. Our approach takes as input two LambdaPix expressions and uses them to output a logic formula. In Section \ref{section:soundness}, we show that if the output formula is valid by Definition \ref{def:valid}, then the two expressions are necessarily equivalent. To define our approach's method of constructing the logic formula from the original LambdaPix expressions in Section \ref{section:Formula Generation}, we begin by first defining a few helper judgments pertaining to weak head reduction and freshening.

\subsection{Weak Head Reduction $e \whnfbig e'$}

We do not have the option of fully evaluating the expressions
during execution, as expressions may contain free
variables in redex positions.
For this reason we use weak head reduction at each step;
this eliminates head-position redexes until free variables get in the way.
The result is a weak head normal form expression.

\newcounter{bigwhnfrule}
\newcommand{\definebigwhnfrule}[1]{\refstepcounter{bigwhnfrule}\label{bigwhnf:#1}\textsc{BigWhnf}_{\thebigwhnfrule}}
\newcommand{\bigwhnfrule}[1]{\ensuremath{\textsc{BigWhnf}_\text{\ref{bigwhnf:#1}}}}

\begin{figure}[h]
{\small
\centering
\begin{mathpar}
\infer[\definebigwhnfrule{step}]{e \whnfbig e''}{
  e \whnfstep e' &
  e' \whnfbig e''
}

\infer[\definebigwhnfrule{val}]{e \whnfbig e}{e \not \whnfstep}
\newcounter{whnfrule}
\newcommand{\definewhnfrule}[1]{\refstepcounter{whnfrule}\label{whnf:#1}\textsc{Whnf}_{\thewhnfrule}}
\newcommand{\whnfrule}[1]{\ensuremath{\textsc{Whnf}_\text{\ref{whnf:#1}}}}

\infer[\definewhnfrule{application}]{e_1\ e_2 \whnfstep e_1'\ e_2}{
  e_1 \whnfstep e_1'
}

\infer[\definewhnfrule{beta}]{
  (\lambda x . e_1) e_2 \whnfstep [e_2/x]e_1
}{}

\infer[\definewhnfrule{projection1}]{
  e \cdot \ell_i \whnfstep e' \cdot \ell_i
}{
  e \whnfstep e'
}

\infer[\definewhnfrule{projection2}]{
  \{\ldots, \ell_i = e, \ldots\} \cdot \ell_i \whnfstep e
}{}
\end{mathpar}
}
\caption{Weak Head Reduction}
\label{fig:weakhead}
\end{figure}

\newcommand{\freshen}[4]{\mathsf{freshen}\ #1.#2 \hookrightarrow #3.#4}
\newcommand{\freshentogether}[5]{\mathsf{FT} (#1, #2) \xhookrightarrow{#5} (#3, #4)}
\newcommand{\equatebindings}[4]{\mathsf{EB} (#1, #2) \hookrightarrow (#3, #4)}

\subsection{Freshening}

It is sometimes useful to generate fresh variables (globally unique variables) to avoid variable
capture.
As single variables are not the only form of binding sites in LamdbaPix,
we generalize this notion to patterns.
When $\freshen{p}{e}{p'}{e'}$, $p'.e'$ is the same as $p.e$ except all
variables bound by $p$ are alpha-varied to fresh variables. The definition of the freshen judgment is given in Figure \ref{fig:freshen}.

\newcounter{freshenrule}
\newcommand{\definefreshenrule}[1]{\refstepcounter{freshenrule}\label{freshen:#1}\textsc{Freshen}_{\thefreshenrule}}
\newcommand{\freshenrule}[1]{\ensuremath{\textsc{Freshen}_\text{\ref{freshen:#1}}}}
\begin{figure}[h]
{\small
\begin{mathpar}
\infer[\definefreshenrule{wildcard}]{\freshen{\_}{e}{\_}{e}}{}

\infer[\definefreshenrule{variable}]{\freshen{x}{e}{y}{[y/x]e}}{
  y \fresh
}

\infer[\definefreshenrule{record}]{
  \freshen{\{\ell_1=p_1, \ldots, \ell_n=p_n\}}{e}
          {\{\ell_1=p_1', \ldots, \ell_n=p_n'\}}{e_n}
}{
  \freshen{p_1}{e}{p_1'}{e_1} &
  \freshen{p_2}{e_1}{p_2'}{e_2} &
  \ldots &
  \freshen{p_n}{e_{n-1}}{p_n'}{e_n}
}

\infer[\definefreshenrule{alias}]{
  \freshen{x\ \mathsf{as}\ p}{e}{y\ \mathsf{as}\ p'}{[y/x]e'}
}{
  y \fresh &
  \freshen{p}{e}{p'}{e'}
}

\infer[\definefreshenrule{constant}]{
  \freshen{c}{e}{c}{e}
}{}

\infer[\definefreshenrule{injection (without arg)}]
{\freshen{i}{e}{i}{e}}{}

\infer[\definefreshenrule{injection (with arg)}]{
  \freshen{i \cdot p}{e}{i \cdot p'}{e'}
}{
  \freshen{p}{e}{p'}{e'}
}
\end{mathpar}
}
\caption{Freshening}
\label{fig:freshen}
\end{figure}

In addition to creating fresh variables to avoid variable capture, our approach sometimes generates fresh variables in order to couple the binding sites between two expressions being considered. For instance, if our approach knows that the same expression $e$ is being matched to variable $x$ in one expression and variable $y$ in another expression, it is useful to equate these bindings so that as our approach proceeds, it is able to know that $x$ in the first expression is the same as $y$ in the second expression. The judgment $\equatebindings{p_1.e_1}{p_2.e_2}{p.e_1'}{p.e_2'}$ defined in Figure $\ref{fig:equatebindings}$ does exactly that, taking in two bindings and returning freshened versions of those bindings that use the same variables so long as the two bindings $p_1.e_1$ and $p_2.e_2$ can be alpha-varied to use a shared pattern $p$.

\newcounter{equatebindingsrule}
\newcommand{\defineequatebindingsrule}[1]{\refstepcounter{equatebindingsrule}\label{equatebindings:#1}\textsc{EB}_{\theequatebindingsrule}}
\newcommand{\equatebindingsrule}[1]{\ensuremath{\textsc{EB}_\text{\ref{equatebindings:#1}}}}
\begin{figure}[!t]
{\small
\begin{mathpar}
\infer[\defineequatebindingsrule{wildcard/wildcard}]
{\equatebindings{\_.e_1}{\_.e_2}{\_.e_1}{\_.e_2}}
{}

\infer[\defineequatebindingsrule{wildcard/variable}]
{\equatebindings{\_.e_1}{x.e_2}{y.e_1}{y.[y/x]e_2}}
{y \fresh}

\infer[\defineequatebindingsrule{variable/wildcard}]
{\equatebindings{x.e_1}{\_.e_2}{y.[y/x]e_1}{y.e_2}}
{y \fresh}

\infer[\defineequatebindingsrule{variable/variable}]
{\equatebindings{x.e_1}{x'.e_2}{y.[y/x]e_1}{y.[y/x']e_2}}
{y \fresh}

\infer[\defineequatebindingsrule{as/other}]
{\equatebindings{x\ \texttt{as}\ p_1.e_1}{p_2.e_2}{y\ \texttt{as}\ p'.[y/x]e_1'}{y\ \texttt{as}\ p'.e_2'}
}
{y \fresh & \equatebindings{p_1.e_1}{p_2.e_2}{p'.e_1'}{p'.e_2'}
}

\infer[\defineequatebindingsrule{other/as}]
{\equatebindings{p_1.e_1}{x\ \texttt{as}\ p_2.e_2}{y\ \texttt{as}\ p'.e_1'}{y\ \texttt{as}\ p'.[y/x]e_2'}
}
{y \fresh & \equatebindings{p_1.e_1}{p_2.e_2}{p'.e_1'}{p'.e_2'}
}

\infer[\defineequatebindingsrule{record/record}]
{\equatebindings{\{\ell_1 = p_1 .. \ell_n = p_n\}.e_1}{\{\ell_1 = p'_1 .. \ell_n = p'_n\}.e_2}{\{\ell_1 = p''_1 .. \ell_n = p''_n\}.e_1^n}{\{\ell_1 = p''_1 .. \ell_n = p''_n\}.e_2^n}
}
{\equatebindings{p_1.e_1}{p_1'.e_2}{p_1''.e_1^1}{p_1''.e_2^1} & \ldots &
\equatebindings{p_n.e_1^{n-1}}{p_n'.e_2^{n-1}}{p_n''.e_1^n}{p_n''.e_2^n}
}

\infer[\defineequatebindingsrule{constant/constant}]
{\equatebindings{c.e_1}{c.e_2}{c.e_1}{c.e_2}}
{}

\infer[\defineequatebindingsrule{injection/injection}]
{\equatebindings{i.e_1}{i.e_2}{i.e_1}{i.e_2}}
{}

\infer[\defineequatebindingsrule{injection/injection2}]
{\equatebindings{i \cdot p_1.e_1}{i \cdot p_2.e_2}{i \cdot p'.e'_1}{i \cdot p'.e'_2}}
{\equatebindings{p_1.e_1}{p_2.e_2}{p'.e'_1}{p'.e'_2}
}
\end{mathpar}
}
\caption{Equate Bindings Judgment}
\label{fig:equatebindings}
\end{figure}

\newcounter{freshentogetherrule}
\newcommand{\definefreshentogetherrule}[1]{\refstepcounter{freshentogetherrule}\label{freshentogether:#1}\textsc{FT}_{\thefreshentogetherrule}}
\newcommand{\freshentogetherrule}[1]{\ensuremath{\textsc{FT}_\text{\ref{freshentogether:#1}}}}
\begin{figure}[h]
{\small
\begin{mathpar}
\infer[\definefreshentogetherrule{singleton}]
{\freshentogether{\{p_1.e_1 \mid \cdot\}}{\{p_2.e_2 \mid \cdot\}}{\{p.e'_1 \mid \cdot\}}{\{p.e'_2 \mid \cdot\}}{1}
}
{\equatebindings{p_1.e_1}{p_2.e_2}{p.e'_1}{p.e'_2}}

\infer[\definefreshentogetherrule{success}]
{\freshentogether{\{p_1.e_1 \mid rest_1\}}{\{p_2.e_2 \mid rest_2\}}{\{p.e'_1 \mid rest'_1\}}{\{p.e'_2 \mid rest'_2\}}{n+1}
}
{\equatebindings{p_1.e_1}{p_2.e_2}{p.e'_1}{p.e'_2} & \freshentogether{rest_1}{rest_2}{rest'_1}{rest'_2}{n}
}

\infer[\definefreshentogetherrule{failure}]
{\freshentogether{\{p_1.e_1 \mid \ldots \mid p_n.e_n\}}{\{p'_1.e'_1 \mid \ldots \mid p'_m.e'_m\}}{\{p_1''.e_1'' \mid \ldots \mid p_n''.e_n''\}}{\{p'''_1.e'''_1 \mid \ldots \mid p'''_m.e'''_m\}}{0}
}
{\forall_{i \in [n]} 
(
\freshen{p_i}{e_i}{p''_i}{e''_i}
)
&
\forall_{i \in [m]}
(
\freshen{p'_i}{e'_i}{p'''_i}{e'''_i}
)
}
\end{mathpar}
}
\caption{Freshen Together Judgment}
\label{fig:freshentogether}
\end{figure}

The benefit of the equate bindings judgment specifically comes into play when comparing case expressions. If two case expressions are casing on the same $e$, and they have identical or near identical binding structures, then it is sometimes useful to freshen the case expressions together, so that as our approach proceeds to consider all of the possible outcomes of the case expressions, it is able to know that the same $e$ was bound in the same way in both expressions. The judgment $\freshentogether{\{p_1.e_1 \mid \ldots \mid p_n.e_n\}}{\{p'_1.e'_1 \mid \ldots \mid p'_m.e'_m\}}{\{p_1''.e_1'' \mid \ldots \mid p_n''.e_n''\}}{\{p_1'''.e_1''' \mid \ldots \mid p_n'''.e_n'''\}}{s}$ defined in Figure \ref{fig:freshentogether} takes in two lists of bindings from case expressions, and equates the first $s$ bindings, independently freshening the rest. The judgment is defined so that once a pair of bindings cannot be equated, all subsequent bindings are freshened independently. This is done to ensure that no bindings are unsoundly equated. The rules listed in Figure \ref{fig:freshentogether} are listed in order of precedence (i.e. if it is possible to apply \freshentogetherrule{success} or \freshentogetherrule{failure}, it will apply \freshentogetherrule{success}).

\subsection{Formula Generation $\Gamma \vdash e_1 \smtequivnormal{\sigma} e_2 : \tau \dashv \Gamma'$}\label{section:Formula Generation}

The judgment that connects the validity of logic formulas with the equivalence of LambdaPix expressions is $\Gamma \vdash e_1 \smtequiv{\sigma} e_2 : \tau \dashv \Gamma'$. The judgment that defines how our approach generates said logic formulas is $\Gamma \vdash e_1 \smtequivnormal{\sigma} e_2 :\tau \dashv \Gamma'$. 

When $\Gamma \vdash e_1 \smtequiv{\sigma} e_2 : \tau \dashv \Gamma'$ or $\Gamma \vdash e_1 \smtequivnormal{\sigma} e_2 : \tau \dashv \Gamma'$, the only free variables appearing in $e_1$ and $e_2$ are in $\Gamma$, so $\Gamma \vdash e_1 : \tau$ and $\Gamma \vdash e_2 : \tau$. However, $\sigma$ can contain more free variables than just those in
$\Gamma$. The purpose of $\Gamma'$ is to describe the rest of the variables in $\sigma$.
$\Gamma$ and $\Gamma'$ are disjoint and between them account for all variables
which may appear in $\sigma$.

\newcounter{isoexprule}
\renewcommand{\theisoexprule}{\textsc{IsoExp}}
\begin{figure}[h]
{\small
\centering
\begin{mathpar}
\infer[\refstepcounter{isoexprule}\label{isoexp}\textsc{IsoExp}]{
  \Gamma \vdash e_1 \smtequiv{\sigma} e_2 : \tau \dashv \Gamma'
}{
  e_1 \whnfbig e'_1 &
  e_2 \whnfbig e'_2 &
  \Gamma \vdash e'_1 \smtequivnormal{\sigma} e'_2 : \tau \dashv \Gamma';
}
\end{mathpar}
}
\caption{IsoExp Rule}
\label{fig:isoexp}
\end{figure}

The judgment $\Gamma \vdash e_1 \smtequiv{\sigma} e_2 : \tau \dashv \Gamma'$ is defined by Figure \ref{fig:isoexp} and is mutually recursive with $\Gamma \vdash e_1 \smtequivnormal{\sigma} e_2 : \tau \dashv \Gamma'$. We use it to define what it means for two expressions to be isomorphic.

\begin{definition}[Isomorphic]
We call two expressions $e_1$ and $e_2$ where $\Gamma_\textnormal{initial} \vdash e_1 : \tau$ and
$\Gamma_\textnormal{initial} \vdash e_2 : \tau$ isomorphic if
$\Gamma_\textnormal{initial} \vdash e_1 \smtequiv{\sigma} e_2 : \tau \dashv \Gamma'$ and
$\forallval_{\Gamma'}. \sigma$.
\end{definition}

The purpose of the distinction between the two judgments is to allow our approach to perform weak head reduction exactly when needed. The judgment $\Gamma \vdash e_1 \smtequivnormal{\sigma} e_2 : \tau \dashv \Gamma'$ assumes as a precondition that $e_1$ and $e_2$ are in weak head normal form, and is defined by Figures \ref{fig:formula generation}, \ref{fig:formula generation app}, and \ref{fig:formula generation case}. 

\newcounter{isorule}
\newcommand{\defineisorule}[1]{
\refstepcounter{isorule}
\label{iso:#1}
\ifnum\pdfstrcmp{#1}{application}=0
\textsc{Iso}_\text{application1}
\else\ifnum\pdfstrcmp{#1}{application var/e}=0 \textsc{Iso}_\text{application2}
\else\ifnum\pdfstrcmp{#1}{application e/var}=0 \textsc{Iso}_\text{application3}
\else\ifnum\pdfstrcmp{#1}{application f/e}=0 \textsc{Iso}_\text{application4}
\else\ifnum\pdfstrcmp{#1}{application e/f}=0 \textsc{Iso}_\text{application5}
\else\ifnum\pdfstrcmp{#1}{application fix/e}=0 \textsc{Iso}_\text{application6}
\else\ifnum\pdfstrcmp{#1}{application e/fix}=0 \textsc{Iso}_\text{application7}
\else\ifnum\pdfstrcmp{#1}{caser}=0 \textsc{Iso}_\text{case2}
\else\ifnum\pdfstrcmp{#1}{symcase}=0 \textsc{Iso}_\text{case3}
\else\ifnum\pdfstrcmp{#1}{freshentogethercase1}=0 \textsc{Iso}_\text{case4}
\else\ifnum\pdfstrcmp{#1}{freshentogethercase2}=0 \textsc{Iso}_\text{case5}
\else
\textsc{Iso}_\text{#1}
\fi\fi\fi\fi\fi\fi\fi\fi\fi\fi\fi
}
\newcommand{\isorule}[1]{
\ifnum\pdfstrcmp{#1}{application}=0
\ensuremath{\textsc{Iso}_\text{application1}}
\else\ifnum\pdfstrcmp{#1}{application var/e}=0 
\ensuremath{\textsc{Iso}_\text{application2}}
\else\ifnum\pdfstrcmp{#1}{application e/var}=0 
\ensuremath{\textsc{Iso}_\text{application3}}
\else\ifnum\pdfstrcmp{#1}{application f/e}=0 \ensuremath{\textsc{Iso}_\text{application4}}
\else\ifnum\pdfstrcmp{#1}{application e/f}=0 \ensuremath{\textsc{Iso}_\text{application5}}
\else\ifnum\pdfstrcmp{#1}{application fix/e}=0 
\ensuremath{\textsc{Iso}_\text{application6}}
\else\ifnum\pdfstrcmp{#1}{application e/fix}=0 
\ensuremath{\textsc{Iso}_\text{application7}}
\else\ifnum\pdfstrcmp{#1}{caser}=0 \ensuremath{\textsc{Iso}_\text{case2}}
\else\ifnum\pdfstrcmp{#1}{symcase}=0 \ensuremath{\textsc{Iso}_\text{case3}}
\else\ifnum\pdfstrcmp{#1}{freshentogethercase1}=0 \ensuremath{\textsc{Iso}_\text{case4}}
\else\ifnum\pdfstrcmp{#1}{freshentogethercase2}=0 \ensuremath{\textsc{Iso}_\text{case5}}
\else
\ensuremath{\textsc{Iso}_\text{#1}}
\fi\fi\fi\fi\fi\fi\fi\fi\fi\fi\fi
}

\begin{figure}[h]
{\small
\begin{mathpar}
\infer[\defineisorule{atomic}]{
  \Gamma \vdash e_1 \smtequivnormal{e_1 \equiv e_2} e_2 : \tau \dashv \cdot
}{
  \Gamma \vdash e_1 : \tau &
  \Gamma \vdash e_2 : \tau &
  e_1 \term &
  e_2 \term
}

\infer[\defineisorule{record}]{
  \Gamma \vdash
    \{\ell_1 = e_1, \ldots, \ell_n = e_n\}
    \smtequivnormal{\sigma_1 \land \ldots \land \sigma_n}
    \{ \ell_1 = e'_1, \ldots, \ell_n = e'_n\}
    : \{ \ell_1 : \tau_1, \ldots, \ell_n : \tau_n \}
  \dashv \Gamma'_1, \ldots, \Gamma'_n
}{
  \Gamma \vdash e_1 \smtequiv{\sigma_1} e'_1 : \tau_1 \dashv \Gamma'_1 &
  \ldots &
  \Gamma \vdash e_n \smtequiv{\sigma_n} e'_n : \tau_n \dashv \Gamma'_n
}

\infer[\defineisorule{projection}]{
  \Gamma \vdash
    e_1 \cdot \ell_i
    \smtequivnormal{\sigma}
    e_2 \cdot \ell_i
    : \tau_i
  \dashv \Gamma'
}{
  \Gamma \vdash
    e_1 \smtequivnormal{\sigma} e_2
    : \{\ldots, \ell_i : \tau_i, \ldots \}
  \dashv \Gamma'
}

\infer[\defineisorule{injection}]{
  \Gamma \vdash
    i \cdot e_1 \smtequivnormal{\sigma} i \cdot e_2
  : \delta \dashv \Gamma'
}{
  i : \tau \hookrightarrow \delta &
  \Gamma \vdash e_1 \smtequiv{\sigma} e_2 : \tau \dashv \Gamma'
}

\infer[\defineisorule{lambda}]{
  \Gamma \vdash
    \lambda x_1 . e_1 \smtequivnormal{\sigma} \lambda x_2 . e_2
    : \tau \rightarrow \tau'
  \dashv x : \tau, \Gamma'
}{
  x \fresh &
  \Gamma, x : \tau \vdash
    [x/x_1]e_1 \smtequiv{\sigma} [x/x_2]e_2
    : \tau'
  \dashv \Gamma'
}

\infer[\defineisorule{fix}]{
  \Gamma \vdash
    \texttt{fix}\ x_1\ \texttt{is}\ e_1
    \smtequivnormal{\sigma}
    \texttt{fix}\ x_2\ \texttt{is}\ e_2
    : \tau
  \dashv x : \tau, \Gamma'
}{
  x \fresh &
  \Gamma, x : \tau \vdash [x/x_1]e_1 \smtequiv{\sigma} [x/x_2]e_2 : \tau \dashv \Gamma'
}
\end{mathpar}
}
\vspace{2mm}
\caption{Formula Generation Rules}
\label{fig:formula generation}
\vspace{2mm}
\end{figure}

Each rule in Figure \ref{fig:formula generation} is written to address a particular syntactic form that $e_1$ and $e_2$ might take. Since each rule targets a particular syntactic form, the premises of each rule are motivated by the semantics of that form. For example, \isorule{lambda} has the premises $x \fresh$ and $\Gamma, x : \tau \vdash [x/x_1]e_1 \smtequiv{\sigma} [x/x_2]e_2 : \tau' \dashv \Gamma'$. The former premise simply declares $x$ as a previously unused variable, and the latter premise states that if any value $x$ of type $\tau$ (the input type to both expressions) is substituted for $x_1$ in the left expression and $x_2$ in the right expression, then the two expressions will be equivalent if $\sigma$ is valid. This reflects the fact that two functions are equivalent if and only if their outputs are equivalent for all valid inputs. 

Although the soundness of these rules is guaranteed, their completeness is not. For instance, \isorule{projection} has the premise $\Gamma \vdash e_1 \smtequivnormal{\sigma} e_2 : \{\ldots, \ell_i : \tau_i, \ldots \} \dashv \Gamma'$. If this premise holds, then the conclusion that $\Gamma \vdash e_1 \cdot \ell_i \smtequivnormal{\sigma} e_2 \cdot \ell_i : \tau_i \dashv \Gamma'$ necessarily follows, as if two records are equivalent, then each of the records' respective entries must also be equivalent. But it is not the case that in order for two projections to be equivalent, they must project from equivalent records. 

\begin{figure}[h]
{\small
\begin{mathpar}
\infer[\defineisorule{application}]{
  \Gamma \vdash
    e_1\ e'_1 \smtequivnormal{\sigma \land \sigma'} e_2\ e'_2
    : \tau'
  \dashv \Gamma', \Gamma''
}{
  \Gamma \vdash e_1 \smtequivnormal{\sigma} e_2 : \tau \rightarrow \tau' \dashv \Gamma' &
  \Gamma \vdash e'_1 \smtequiv{\sigma'} e'_2 : \tau \dashv \Gamma''
}

\infer[\defineisorule{application var/e}]{
  \Gamma \vdash
    x\ e_1 \smtequivnormal{\sigma} e_2
    : \tau
  \dashv \Gamma'
}
{y \fresh &
\Gamma, y : \tau \vdash
    y \smtequivnormal{\sigma} [y/(x\ e_1)]e_2
    : \tau
  \dashv \Gamma'
}

\infer[\defineisorule{application e/var}]{
  \Gamma \vdash
    e_1 \smtequivnormal{\sigma} x\ e_2
    : \tau
  \dashv \Gamma'
}
{y \fresh &
\Gamma, y : \tau \vdash
    [y/(x\ e_2)]e_1 \smtequivnormal{\sigma} y
    : \tau
  \dashv \Gamma'
}

\infer[\defineisorule{application f/e}]{
  \Gamma \vdash
    o\ e_1 \smtequivnormal{\sigma} e_2
    : \tau
  \dashv \Gamma'
}
{y \fresh &
\Gamma, y : \tau \vdash
    y \smtequivnormal{\sigma} [y/(o\ e_1)]e_2
    : \tau
  \dashv \Gamma'
}

\infer[\defineisorule{application e/f}]{
  \Gamma \vdash
    e_1 \smtequivnormal{\sigma} o\ e_2
    : \tau
  \dashv \Gamma'
}
{y \fresh &
\Gamma, y : \tau \vdash
    [y/(o\ e_2)]e_1 \smtequivnormal{\sigma} y
    : \tau
  \dashv \Gamma'
}

\infer[\defineisorule{application fix/e}]{
  \Gamma \vdash
    (\texttt{fix}\ x_1\ \texttt{is}\ e_1)\ e_2 \smtequivnormal{\sigma} e
    : \tau
  \dashv \Gamma'
}
{y \fresh &
\Gamma, y : \tau \vdash
    y \smtequivnormal{\sigma} [y/((\texttt{fix}\ x_1\ \texttt{is}\ e_1)\ e_2)] e
    : \tau
  \dashv \Gamma'
}

\infer[\defineisorule{application e/fix}]{
  \Gamma \vdash
    e \smtequivnormal{\sigma} (\texttt{fix}\ x_1\ \texttt{is}\ e_1)\ e_2
    : \tau
  \dashv \Gamma'
}
{y \fresh &
\Gamma, y : \tau \vdash
    [y/((\texttt{fix}\ x_1\ \texttt{is}\ e_1)\ e_2)] e \smtequivnormal{\sigma} y
    : \tau
  \dashv \Gamma'
}
\end{mathpar}
}
\vspace{2mm}
\caption{Formula Generation Application Rules}
\label{fig:formula generation app}
\vspace{2mm}
\end{figure}

Each rule in Figure \ref{fig:formula generation app} addresses the case in which at least one of the expressions being compared is an application. When the two expressions being compared are both applications of equivalent arguments onto equivalent functions, \isorule{application} can be used to infer equivalence of the resulting applications. For situations in which an application is being compared to another syntactic form, or two applications that cannot be recognized as equivalent via \isorule{application} are being compared, the remaining rules take an application and replace it with a shared fresh variable in both expressions. For example, if the expressions $f(x)$ and $f(x+0)$ are being compared, \isorule{application} is sufficient to find equivalence because $f$ can be found equivalent to $f$ and $x$ can be found equivalent to $x+0$ via \isorule{atomic}. But if $f(x)$ and $f(x)+0$ are being compared, \isorule{application} alone would be insufficient, as the outermost function of the first expression is $f$ and the outermost function of the second expression is $+$. For this situation, \isorule{application var/e} is needed to replace $f(x)$ with the fresh variable $y$, yielding the expressions $y$ and $y+0$, which can be immediately found equivalent via \isorule{atomic}.

The current formula generation application rules have multiple limitations. First, the rules only allow applications to be replaced with shared fresh variables when the application being replaced is at the outermost level of one of the expressions. This has the consequence that although $f(x) + f(x)$ and $2 * f(x)$ are obviously equivalent, and the substitution of $f(x)$ for a shared fresh variable $y$ would enable \isorule{atomic} to prove that fact, our current rules do not support this inference. Second, \isorule{application fix/e} and \isorule{application e/fix} require substituting an entire fixed point application in an expression, so unless if the two expressions being compared have essentially identical fixed points included, these rules will be ineffective. Still, despite these limitations, the current formula generation application rules are sufficient for their most common purpose of working with \isorule{fix} to ensure that recursive function calls are recognized as equivalent when given equivalent arguments.

\begin{figure}[h]
{\small
\begin{mathpar}
\infer[\defineisorule{casel}]{
  \Gamma \vdash
    \texttt{case}\ e\ \{ p_1 . e_1 \mid \ldots \mid p_n . e_n \} \smtequivnormal{\land_{i \in [n]} ((\land_{j \in [i-1]} (e \not\equiv p_j')) \land e \equiv p_i') \Rightarrow \sigma_i} e'
    : \tau
  \dashv \forall_{i \in [n]} \Gamma_i, \Gamma_i'
}{
  e \term &
  \forall_{i \in [n]} \left(
    \freshen{p_i}{e_i}{p_i'}{e_i'} \quad
    p_i' :: \tau' \dashv \Gamma_i \quad
    \Gamma, \Gamma_i \vdash
      e_i' \smtequiv{\sigma_i} e' : \tau \dashv \Gamma'_i
  \right)
}

\infer[\defineisorule{caser}]{
  \Gamma \vdash
    e' \smtequivnormal{\land_{i \in [n]} ((\land_{j \in [i-1]}(e \not\equiv p_j')) \land e \equiv p_i') \Rightarrow \sigma_i}
    \texttt{case}\ e\ \{ p_1 . e_1 \mid \ldots \mid p_n . e_n \} 
    : \tau
  \dashv \forall_{i \in [n]} \Gamma_i, \Gamma_i'
}{
  e \term &
  \forall_{i \in [n]} \left(
    \freshen{p_i}{e_i}{p_i'}{e_i'} \quad
    p_i' :: \tau' \dashv \Gamma_i \quad
    \Gamma, \Gamma_i \vdash
      e_i' \smtequiv{\sigma_i} e' : \tau \dashv \Gamma'_i
  \right)
}

\infer[\defineisorule{symcase}]{
  \Gamma \vdash 
    \texttt{case}\ e\ \{ \ldots \}
    \smtequivnormal{\sigma \land \sigma'}
    \texttt{case}\ e'\ \{ \ldots' \}
    : \tau
  \dashv \Gamma', x : \tau', \Gamma''
}{
  \Gamma \vdash e \smtequivnormal{\sigma} e' : \tau' \dashv \Gamma' &
  x \fresh &
  \Gamma, x : \tau' \vdash
    \texttt{case}\ x\ \{ \ldots \}
    \smtequivnormal{\sigma'}
    \texttt{case}\ x\ \{ \ldots' \}
    : \tau
  \dashv \Gamma''
}

\inferrule 
{
\freshentogether{\{M\}}{\{M'\}}{\{p_1.e_1 \mid \ldots \mid p_n.e_n\}}{\{p'_1.e'_1 \mid \ldots \mid p'_m.e'_m\}}{s} \\
\forall_{i \in [s]}
(
p_i :: \tau' \dashv \Gamma_i \quad
\Gamma, \Gamma_i \vdash e_i \smtequiv{\sigma_i} e'_i \dashv \Gamma_i'
) \\\\
\forall_{j \in [s+1, n]}
(p_j :: \tau' \dashv \Gamma_j \quad \Gamma, \Gamma_j \vdash e_j \smtequiv{\sigma_j} \texttt{case}\ x\ \{p'_1.e'_1 \mid \ldots \mid p'_m.e'_m\} : \tau \dashv \Gamma'_j
)
}
{
  \Gamma \vdash 
    \texttt{case}\ x\ \{M\}
    \smtequivnormal{\Psi}
    \texttt{case}\ x\ \{M'\}
    : \tau
  \dashv \forall_{i \in [n]}\Gamma_i, \Gamma'_i
}
\defineisorule{freshentogethercase1}

\inferrule 
{
\freshentogether{\{M'\}}{\{M\}}{\{p'_1.e'_1 \mid \ldots \mid p'_m.e'_m\}}{\{p_1.e_1 \mid \ldots \mid p_n.e_n\}}{s} \\
\forall_{i \in [s]}
(
p_i :: \tau' \dashv \Gamma_i \quad
\Gamma, \Gamma_i \vdash e_i \smtequiv{\sigma_i} e'_i \dashv \Gamma_i'
) \\\\
\forall_{j \in [s+1, n]}
(p_j :: \tau' \dashv \Gamma_j \quad \Gamma, \Gamma_j \vdash \texttt{case}\ x\ \{p'_1.e'_1 \mid \ldots \mid p'_m.e'_m\} \smtequiv{\sigma_j} e_j : \tau \dashv \Gamma'_j
)
}
{
  \Gamma \vdash 
    \texttt{case}\ x\ \{M'\}
    \smtequivnormal{\Psi}
    \texttt{case}\ x\ \{M\}
    : \tau
  \dashv \forall_{i \in [n]}\Gamma_i, \Gamma'_i
}
\defineisorule{freshentogethercase2}

$$\Psi := (\land_{i \in [s]}\sigma_i) \land (\land_{j \in [s+1, n]} ((\land_{k \in [j-1]} (x \not\equiv p_k)) \land x \equiv p_j) \Rightarrow \sigma_j)$$
\end{mathpar}
}
\vspace{2mm}
\caption{Formula Generation Case Rules}
\label{fig:formula generation case}
\vspace{2mm}
\end{figure}

Each rule in Figure \ref{fig:formula generation case} addresses the situation in which at least one of the expressions being compared is a case analysis. These rules can be grouped into two broad approaches. For situations in which only one of the expressions being compared is a case analysis, or both expressions are case analyses but the expressions being cased on are not equivalent, \isorule{casel} and \isorule{caser} are used to unpack one case analysis at a time. If $\texttt{case}\ e\ \{ p_1 . e_1 \mid \ldots \mid p_n . e_n \}$ is being compared to $e'$, then the formula generated by these rules states that if $e$ can be pattern matched with $p_i$ and no prior patterns, $e_i$ needs to be equivalent to $e'$ in order for the two overall expressions to be equivalent. 

For situations in which the two expressions being compared are case analyses that are casing on equivalent expressions, \isorule{symcase}, \isorule{freshentogethercase1}, and \isorule{freshentogethercase2} are used to deconstruct both case expressions simultaneously. To do this, \isorule{symcase} is always used first to ensure that the expressions being cased on are equivalent. If the expressions being cased on are not equivalent, then $\sigma$ in the formula generated by \isorule{symcase} will not be valid, and so the output formula $\sigma \land \sigma'$ will not be valid as a result. If the expressions being cased on are equivalent, then \isorule{freshentogethercase1} and \isorule{freshentogethercase2} can be used to generate $\sigma'$. This approach is needed in addition to \isorule{casel} and \isorule{caser} because \isorule{casel} and \isorule{caser} require that the expression being cased on is a base term.

All rules in Figure \ref{fig:formula generation} are deterministic in the sense that for all possible expressions, at most one rule is applicable. However, the rules in Figures \ref{fig:formula generation app} and \ref{fig:formula generation case} are non-deterministic. If two case expressions or two applications are being compared, there may be multiple applicable rules. For instance, if $\texttt{case}\ 1\ \{1.2 | \_.3\}$ is being compared to $\texttt{case}\ 2\ \{\_.2\}$, then \isorule{casel}, \isorule{caser}, and \isorule{symcase} are all applicable. Our approach handles this by considering all formulas that can be generated by applying any applicable rule and outputs the disjunction of all generated formulas. We will later show that applying any applicable rule in such a situation is sound and that therefore, taking the disjunction of all generated formulas is also sound. The only exception to this is that \isorule{symcase} cannot be applied multiple times in a row because it is never useful to do so and allowing this would cause an infinite loop.

In instances where there is no applicable rule, such as if $i_1\ e_1$ is compared with $i_2\ e_2$ where $i_1 \neq i_2$ and either $e_1$ or $e_2$ cannot be encoded into a term, our approach simply outputs the formula $\sigma$=False, which is always sound.

\subsection{Limitations and Further Extensions}

The current set of rules is comprehensive and covers a wide range of operators that are often found in many functional programming assignments. However, there are limitations to the current set of rules, some of which have already been noted. The current main limitations include:

\begin{itemize}
    \item Our current handling of projections in \isorule{projection} requires that in order for two projections to be recognized as equivalent, they must project from equivalent records.
    \item Our current handling of recursive function calls occurs entirely through the interplay between \isorule{fix} and the formula generation application rules. Because of how these rules are currently defined, recursive functions can only be recognized as equivalent if in all situations they recurse on equivalent arguments or do not recurse at all.
    \item The approach taken by the formula generation application rules is limited in that applications can only be replaced with shared fresh variables when the application being replaced is at the outermost level of one of the expressions being compared.
    \item \isorule{application fix/e} and \isorule{application e/fix} both require substituting a variable for an entire fixed point application, which will only be useful if the two expressions being compared have essentially identical fixed points included.
    \item Since \isorule{casel} and \isorule{caser} require that the expression being cased on is a base term, the current set of rules cannot identify equivalence between a case analysis in which the expression being cased on isn't a base term and any other syntactic form.
    \item The current definition of LambdaPix does not allow for state, and so our approach cannot identify the equivalence of any programs that use state.
\end{itemize}

Compared to other potential extensions that could be implemented to address an aforementioned limitation, extending LambdaPix to support state would likely require a significant number of changes to our approach. However, this could be potentially achieved by handling sequential state-altering declaration similar to how we handle local declaration. Currently, we handle local declaration by encoding the declaration into the SMT formula in the same way that we would encode a single pattern case expression (i.e. $\texttt{let\ val\ x\ =}\ e_1\ \texttt{in}\ e_2\ \texttt{end}$ becomes $\texttt{case}\ e_1\ \{x.e_2\}$ at the transpilation to LambdaPix stage). It would not be possible to do the same procedure for sequential declaration since the scoping would have to be global. However, we believe that it may be feasible to treat reference declaration/assignment similar to variable declaration/initialization and reference update similar to variable shadowing with modified scoping.

One advantage of the structure of our approach is that extending our system to address some of the previously listed limitations is straightforward. As soon as a new rule that addresses one of the system's current limitations is found to be sound, it can be simply tacked on to the current system without needing to modify any preexisting rules. This also applies to extensions of the underlying language LambdaPix itself. Adding new base types to LambdaPix such as strings or reals requires no modification of the current rules whatsoever, and adding additional syntactic expression forms requires only the addition of rules for comparing the new form against itself and arbitrary expressions. Even though it is easy to extend the LambdaPix language and add additional rules, the current version is already rich enough to capture common behavior in programming assignments of introductory courses.

\section{Operation}\label{section:operation}

To provide a better understanding of our approach, we step through our approach's operation on a pair of simple Standard ML expressions provided above. As we step through this example, we will refer to the inference rules from
the previous section to illustrate how they are applied.

{\centering
\begin{minipage}[t]{0.45\textwidth}
\vspace{0pt}
\centering
\begin{tabular}{c}
\begin{lstlisting}
fun add_opt x y =
  case (x, y) of
    (SOME m, SOME n) =>
      SOME (m + n)
  | (NONE, _) => NONE
  | (_, NONE) => NONE
\end{lstlisting}
\end{tabular}
\end{minipage}
\begin{minipage}[t]{0.45\textwidth}
\vspace{0pt}
\centering
\begin{tabular}{c}
\begin{lstlisting}
fun bind a f =
  case a of
    SOME b => f b
  | NONE => NONE

val return = SOME

fun add_opt x y =
  bind x (fn m =>
  bind y (fn n =>
    return (m + n)
  ))
\end{lstlisting}
\end{tabular}
\end{minipage}}\\

First, we transpile both expressions to LambdaPix. This is shown above. Since much of the proof derivation which drives our approach is free of
branching, through most of this section we will view our approach as
transforming the above expressions through the application of rules, rather than
building up a proof tree.

\begin{minipage}[t]{0.45\textwidth}
\vspace{0pt}
\centering
\begin{tabular}{c}
\begin{lstlisting}[escapeinside={/*}{*/}]
/*$\lambda$*/x./*$\lambda$*/y.
  case (x,y) of
  { (SOME/*$\cdot$*/m,SOME/*$\cdot$*/n).SOME/*$\cdot$*/(m+n)
  | (NONE,_).NONE
  | (_,NONE).NONE }
\end{lstlisting}
\end{tabular}
\end{minipage}
\begin{minipage}[t]{0.45\textwidth}
\vspace{0pt}
\centering
\begin{tabular}{c}
\begin{lstlisting}[escapeinside={/*}{*/}]
/*$\lambda$*/x./*$\lambda$*/y.
  (/*$\lambda$*/a./*$\lambda$*/f.
    case a of
    { SOME/*$\cdot$*/b.f b
    | NONE.NONE }
  ) x (/*$\lambda$*/m.
  (/*$\lambda$*/a./*$\lambda$*/f.
    case a of
    { SOME/*$\cdot$*/b.f b
    | NONE.NONE }
  ) y (/*$\lambda$*/n.
    (/*$\lambda$*/e.SOME/*$\cdot$*/e) (m+n)
  ))
\end{lstlisting}
\end{tabular}
\end{minipage}\\

The entry point to our approach is the
$\Gamma \vdash e_1 \smtequiv{\sigma} e_2 : \tau \dashv \Gamma'$
judgement, defined by the rule \ref{isoexp}.
By this rule, we reduce both expressions to weak head normal form then
apply the $\Gamma \vdash e_1 \smtequivnormal{\sigma} e_2 : \tau \dashv \Gamma'$
judgement to them. However, since the expressions in consideration are abstractions, the expressions are already in weak head normal form, so no transformation is necessary to apply this rule.

\begin{minipage}[t]{0.45\textwidth}
\vspace{0pt}
\centering
\begin{tabular}{c}
\begin{lstlisting}[escapeinside={/*}{*/}]
case (x,y) of
{ (SOME/*$\cdot$*/m,SOME/*$\cdot$*/n).SOME/*$\cdot$*/(m+n)
| (NONE,_).NONE
| (_,NONE).NONE }
\end{lstlisting}
\end{tabular}
\end{minipage}
\begin{minipage}[t]{0.45\textwidth}
\vspace{0pt}
\centering
\begin{tabular}{c}
\begin{lstlisting}[escapeinside={/*}{*/}]
  (/*$\lambda$*/a./*$\lambda$*/f.
    case a of
    { SOME/*$\cdot$*/b.f b
    | NONE.NONE }
  ) x (/*$\lambda$*/m.
   (/*$\lambda$*/a./*$\lambda$*/f.
     case a of
     { SOME/*$\cdot$*/b.f b
     | NONE.NONE }
   ) y (/*$\lambda$*/n.
     (/*$\lambda$*/e.SOME/*$\cdot$*/e) (m+n)
  ))
\end{lstlisting}
\end{tabular}
\end{minipage}\\

Next, since both expressions are lambda expressions with two curried arguments,
we proceed with two applications of the rule \isorule{lambda}. This requires us to create two new fresh variables and substitute them for the first two function arguments in both expressions. For simplicity, we will simply call the first fresh variable \texttt{x} and the second fresh variable \texttt{y} even though these names conflict with the original variable names. The key difference between before and after this process is that before this process, the two functions had the same variable names \texttt{x} and \texttt{y} by coincidence, whereas after this process, the two functions use the same fresh variables \texttt{x} and \texttt{y} by design. These applications of \isorule{lambda} yield the above expressions.

\begin{minipage}[t]{0.45\textwidth}
\vspace{0pt}
\centering
\begin{tabular}{c}
\begin{lstlisting}[escapeinside={/*}{*/}]
case (x,y) of
{ (SOME/*$\cdot$*/m,SOME/*$\cdot$*/n).SOME/*$\cdot$*/(m+n)
| (NONE,_).NONE
| (_,NONE).NONE }
\end{lstlisting}
\end{tabular}
\end{minipage}
\begin{minipage}[t]{0.45\textwidth}
\vspace{0pt}
\centering
\begin{tabular}{c}
\begin{lstlisting}[escapeinside={/*}{*/}]
case x of
{ SOME/*$\cdot$*/b. 
   (/*$\lambda$*/m.
    (/*$\lambda$*/a./*$\lambda$*/f.
     case a of
     { SOME/*$\cdot$*/b.f b
     | NONE. NONE }
    ) y (/*$\lambda$*/n.(/*$\lambda$*/e.SOME/*$\cdot$*/e) (m+n))
   ) b
| NONE.NONE }
\end{lstlisting}
\end{tabular}
\end{minipage}\\

Since the premise of \isorule{lambda} invokes the $\expiso$ judgement, \ref{isoexp} requires that we reduce both expressions to weak head normal form. The left expression is already in weak head normal form, so no transformation is necessary, but the right expression must undergo two beta reductions before it is in weak head normal form. The result of these beta reductions is above.

Since both expressions are case expressions, our approach has multiple options for how to proceed. Formally, our approach pursues all of these options, generating separate formulas for each option, and finally outputting a disjunction of all of the generated formulas. This ensures that if any option can generate a valid formula, then the final result will be the disjunction of the valid formula with several other formulas, which altogether is valid. In this case, attempting to proceed with \isorule{symcase} will not yield a valid formula because the two expressions are casing on different things, but applying either \isorule{casel} or \isorule{caser} can yield a valid formula. For this demonstration, we step through the derivation that results from applying \isorule{casel} and call the formulas generated by applying \isorule{caser} or \isorule{symcase} $\sigma_{\isorule{caser}}$ and $\sigma_{\isorule{symcase}}$ respectively.

As there are three branches in the left case expression, our approach's proof tree now splits into three branches. For this demonstration, we just step through the first of these branches, as the other two branches work similarly. We call the formulas generated by the other two branches of the proof tree $\sigma_\textnormal{branch 2}$ and $\sigma_\textnormal{branch 3}$.

\begin{minipage}[t]{0.45\textwidth}
\vspace{0pt}
\centering
\begin{tabular}{c}
\begin{lstlisting}[escapeinside={/*}{*/}]
SOME/*$\cdot$*/(m1+n1)
\end{lstlisting}
\end{tabular}
\end{minipage}
\begin{minipage}[t]{0.45\textwidth}
\vspace{0pt}
\centering
\begin{tabular}{c}
\begin{lstlisting}[escapeinside={/*}{*/}]
case x of
{ SOME/*$\cdot$*/b. 
   (/*$\lambda$*/m.
    (/*$\lambda$*/a./*$\lambda$*/f.
     case a of
     { SOME/*$\cdot$*/b.f b
     | NONE. NONE }
    ) y (/*$\lambda$*/n.(/*$\lambda$*/e.SOME/*$\cdot$*/e) (m+n))
   ) b
| NONE.NONE }
\end{lstlisting}
\end{tabular}
\end{minipage}\\

We "freshen" the branch selected to avoid variable capture.
In this situation we will freshen the first branch of the left case expression by replacing \texttt{m} and \texttt{n} with \texttt{m1} and
\texttt{n1}, respectively.
From this branch we will generate a formula of the form
\[{\small (\texttt{(x,y)}\equiv\texttt{(SOME$\cdot$m1,SOME$\cdot$n1)}) \Rightarrow \ldots}\]
where the ellipses is what we are going to fill in as we complete this branch
of the proof tree.

Since the left expression has been simplified to a base term, our approach proceeds to work on the right expression. Our approach applies \isorule{caser} twice (using beta reduction to reduce the expression to weak head normal form as appropriate), and finishes each branch of the proof tree by using \isorule{atomic} to compare base terms.

Putting everything together, the final formula is:
{\small
\begin{align*}
    (\sigma_{\textnormal{branch 1}} \land \sigma_{\textnormal{branch 2}} \land \sigma_{\textnormal{branch 3}}) \vee \sigma_{\isorule{caser}} \vee \sigma_{\isorule{symcase}}
\end{align*}
}
where $\sigma_{\textnormal{branch 1}}$ is
{\small
\begin{align*}
&(\texttt{(x,y)}\equiv\texttt{(SOME$\cdot$m1,SOME$\cdot$n1)}) \Rightarrow \\
&\qquad ((\texttt{x}\equiv\texttt{SOME$\cdot$b1}) \Rightarrow \\
&\qquad \qquad (\texttt{y}\equiv\texttt{SOME$\cdot$b2}) \Rightarrow
      (\texttt{SOME$\cdot$(m1+n1)}\equiv\texttt{SOME$\cdot$(b1+b2)}) \land \\
&\qquad \qquad (\texttt{y}\not\equiv\texttt{SOME$\cdot$b2}\land \texttt{y}\equiv\texttt{NONE}) \Rightarrow
      (\texttt{SOME$\cdot$(m1+n1)}\equiv\texttt{NONE}) \\
&\qquad ) \land \\
&\qquad ((\texttt{x}\not\equiv\texttt{SOME$\cdot$b1} \land \texttt{x}\equiv\texttt{NONE}) \Rightarrow \\
&\qquad \qquad (\texttt{y}\equiv\texttt{SOME$\cdot$b2}) \Rightarrow
      (\texttt{SOME$\cdot$(m1+n1)}\equiv\texttt{NONE}) \land \\
&\qquad \qquad (\texttt{y}\not\equiv\texttt{SOME$\cdot$b2} \land \texttt{y}\equiv\texttt{NONE}) \Rightarrow
      (\texttt{SOME$\cdot$(m1+n1)}\equiv\texttt{NONE}) \\
&\qquad )
\end{align*}
}
and $\sigma_\textnormal{branch 2}$ and $\sigma_\textnormal{branch 3}$ are similar.

Since the two original expressions were equivalent, this formula is valid. The validity of this formula can be verified either by hand or by an SMT Solver.

\section{Soundness}\label{section:soundness}
We prove the soundness of our approach: if our approach takes in two expressions and outputs a valid formula, then the two expressions must be equivalent.

\subsection{Extensional Equivalence}

To prove the soundness of our approach, we must first define what it means for two expressions to be equivalent.
For this, we introduce extensional equivalence, a widely accepted
notion of equivalence.
Extensional equivalence is the same as contextual
equivalence, and so two extensionally equivalent expressions are indistinguishable in terms of behavior. This implies that extensional equivalence is closed under evaluation.
Extensional equivalence is also an equivalence relation, so we may assume
that it is reflexive, symmetric, and transitive.
LambdaPix enjoys referential transparency, meaning that
extensional equivalence of LambdaPix expressions is closed under
replacement of subexpressions with extensionally equivalent subexpressions.

\newcommand{\eeq}{\cong}
We use $e_1 \eeq e_2 : \tau$ to denote that expressions $e_1$ and $e_2$ are
extensionally equivalent and both have the type $\tau$.

\newcounter{eeqrule}
\newcommand{\defineeeqrule}[1]{\refstepcounter{eeqrule}\label{eeq:#1}\textsc{EQ}_{\theeeqrule}}
\newcommand{\eeqrule}[1]{\ensuremath{\textsc{EQ}_\text{\ref{eeq:#1}}}}
\begin{definition}[Extensional Equivalence]
We define that $e_1 \eeq e_2 : \tau$ if
$\Gamma_{\textnormal{initial}} \vdash e_1 : \tau$,
$\Gamma_{\textnormal{initial}} \vdash e_2 : \tau$,
$\evaluatesto{e_1}{v_1}$,
$\evaluatesto{e_2}{v_2}$, and

\begin{enumerate}
\item Rule $\defineeeqrule{extensionality}$:
  In the case that $\tau = \tau_1 \rightarrow \tau_2$,
  for all expressions $v$ such that $\Gamma_{\textnormal{initial}} \vdash v : \tau_1$,
  $v_1\ v \eeq v_2\ v : \tau_2$.
\item Rule $\defineeeqrule{matching}$:
  In the case that $\tau$ is not an arrow type,
  for all patterns $p$ such that $p :: \tau$, either $\matches{v_1}{p}{B}$ and
  $\matches{v_2}{p}{B}$ or $\nomatch{v_1}{p}$ and $\nomatch{v_2}{p}$.
\end{enumerate}
\end{definition}

Unlike our approach, extensional equivalence inducts over the types of the
expressions rather than their syntax, and is defined only over closed
expressions. As we are only concerned with proving our approach sound over valuable
expressions, we leave extensional equivalence undefined for divergent
expressions.

This is an atypical formalization of extensional equivalence;
it is typically defined in terms of the elimination
forms of each type connective.
However, since pattern matching in LambdaPix subsumes the elimination
of all connectives other than arrows, we simply define equivalence at all
non-arrow types in terms of pattern matching.

The soundness theorem for our approach connects our technique's definition of isomorphic with this definition of extensional equivalence. It is as follows:

\begin{theorem}[Soundness]
For any expressions $e_1$ and $e_2$, if $\Gamma_{\textnormal{initial}} \vdash e_1 \smtequiv{\sigma} e_2 : \tau \dashv \Gamma'$ and
$\forallval_{\Gamma'} . \sigma$, then $e_1 \eeq e_2 : \tau$.
\end{theorem}

\subsection{Proof Sketch}
As the $\Gamma \vdash e_1 \smtequiv{\sigma} e_2 : \tau \dashv \Gamma'$ judgement is
defined simultaneously with the
$\Gamma \vdash e_1 \smtequivnormal{\sigma} e_2 : \tau \dashv \Gamma'$ judgement,
we prove the theorem by simultaneous induction on both of these judgements.
We also use the $\forallval_\Gamma . j$ judgement to strengthen the
inductive hypotheses to account for variables.
Recall that
if $\Gamma = \vec{x} : \vec{\tau}$, then the judgement 
$\forallval_\Gamma . j$ holds if for all $\vec{v}$ where 
$v_i : \tau_i$ and $\val{v_i}$ for all $v_i \in \vec{v}$,
it is the case that
$[\vec{v}/\vec{x}]j$ holds (implicitly, we omit any primitive operations from the context $\Gamma = \vec{x} : \vec{\tau}$ as to not range over all possible meanings for LambdaPix's primitive operations).
The theorem we wish to show by induction is then:
\begin{itemize}
\item
  If $\Gamma \vdash e_1 \smtequiv{\sigma} e_2 : \tau \dashv \Gamma'$ then
  $\forallval_{\Gamma} . \left(
    \text{if $\left(\forallval_{\Gamma'} . \sigma \right)$ then $e_1 \eeq e_2 : \tau$}
  \right)$.
\item
  If $\Gamma \vdash e_1 \smtequivnormal{\sigma} e_2 : \tau \dashv \Gamma'$ then
  $\forallval_{\Gamma} . \left(
    \text{if $\left(\forallval_{\Gamma'} . \sigma \right)$ then $e_1 \eeq e_2 : \tau$}
  \right)$.
\end{itemize}
We first verify that the above statements imply the soundness theorem. Indeed, when $\Gamma_{\textnormal{initial}} \vdash e_1 \smtequiv{\sigma} e_2 : \tau \dashv \Gamma'$
we have $\forallval_{\Gamma_{\textnormal{initial}}} . \left(\text{if $\left(\forallval_{\Gamma'} . \sigma \right)$ then $e_1 \eeq e_2 : \tau$}\right)$. Since $\Gamma_{\textnormal{initial}}$ contains only primitive operations, which are omitted from the $\forallval_\Gamma.j$ judgment, the outer quantifier quantifies over no variables, so we have that
$\text{if $\left(\forallval_{\Gamma'} . \sigma \right)$ then $e_1 \eeq e_2 : \tau$}$.
This together with the assumption that $\forallval_{\Gamma'} . \sigma$
allows us to conclude that $e_1 \eeq e_2 : \tau$.

The full proof of each rule's soundness has 18 cases and uses 14 lemmas and can be found in~\ref{sec:appendix}. Two cases are included below as examples:\\

\noindent \isorule{record}:
  Let $\Gamma = \vec{x} : \vec{\tau}$
  and let $\vec{v}$ be arbitrary where $v_i : \tau_i$ and $\val{v_i}$
  for all $v_i \in \vec{v}$.
  Assume $[\vec{v}/\vec{x}]
    \left(\forallval_{\Gamma'_1, \ldots, \Gamma'_n}.\sigma_1 \land \ldots \sigma_n\right)
  $. It must be shown that $[\vec{v}/\vec{x}](
    \{\ell_1 = e_1, \ldots, \ell_n = e_n\}
    \eeq
    \{ \ell_1 = e'_1, \ldots, \ell_n = e'_n\}
  )$.
  
\begin{lemma}\label{eeqmatchmain}
If $e_1 \eeq e_2 : \tau$,
  $\evaluatesto{e_1}{v_1}$, and
  $\evaluatesto{e_2}{v_2}$,
  then for all patterns $p$ where
  $p :: \tau \dashv \Gamma$,
  it is the case that
  either $\matches{v_1}{p}{B}$ and $\matches{v_2}{p}{B}$ or
  $\nomatch{v_1}{p}$ and $\nomatch{v_2}{p}$.
  \textit{Proof}: by induction on $e_1 \eeq e_2 : \tau$.
  If $\tau = \tau_1 \rightarrow \tau_2$ then by inversion of
  $p :: \tau \dashv \Gamma$, $p$ must either be a wildcard or a variable.
  Then by \matchrule{wildcard} and \matchrule{variable}, we have that
  $\matches{v_1}{p}{B}$ and $\matches{v_2}{p}{B}$.
  If $\tau$ isn't an arrow type, then we conclude by $\eeqrule{matching}$.
\end{lemma}

  By conjunction and that all the $\Gamma'_i$ are disjoint, we have that
  for all $i \in [n]$, $\forallval_{\Gamma'_i} . \sigma_i$.
  Then by the inductive hypotheses, we have that $[\vec{v}/\vec{x}](e_i \eeq e_i' : \tau_i)$.
  Since we are only concerned with proving our approach sound over valuable expressions, without loss of generality, we can assume that $\evaluatesto{[\vec{v}/\vec{x}]e_i}{v_i}$ and $\evaluatesto{[\vec{v}/\vec{x}]e'_i}{v'_i}$ for some values $v_i$ and $v_i'$.
  By Lemma \ref{eeqmatchmain}, we have that for all $p_i$ where $p_i :: \tau_i \dashv \Gamma_i$,
  either $\matches{v_i}{p_i}{B_i}$ and $\matches{v'_i}{p_i}{B_i}$ or $\nomatch{v_i}{p_i}$ and 
  $\nomatch{v'_i}{p_i}$.\\

  To appeal to \eeqrule{matching},
  let $p$ be an arbitrary pattern such that
  $p :: \{\ell_1 : \tau_1, \ldots, \ell_n : \tau_n\} \dashv \Gamma'$.
  We proceed by cases:
  \begin{itemize}
  \item In the case that for all $i \in [n]$
    $\matches{v_i}{p_i}{B_i}$ and $\matches{v'_i}{p_i}{B_i}$,
    by \matchrule{record1} we have
    $\matches{\{\ell_1 = v_1, \ldots, \ell_n = v_n\}}{p}{B_1 \ldots B_n}$ and
    $\matches{\{\ell_1 = v'_1, \ldots, \ell_n = v'_n\}}{p}{B_1 \ldots B_n}$.
  \item In the case that there is some $i \in [n]$ where
    $\nomatch{v_i}{p_i}$ and $\nomatch{v'_i}{p_i}$,
    by \matchrule{record2} we have
    $\nomatch{\{\ell_1 = v_1, \ldots, \ell_n = v_n\}}{p}$ and
    $\nomatch{\{\ell_1 = v'_1, \ldots, \ell_n = v'_n\}}{p}$.
  \end{itemize}
  Since in all cases either
  $\matches{\{\ell_1 = v_1, \ldots, \ell_n = v_n\}}{p}{B}$ and
  $\matches{\{\ell_1 = v'_1, \ldots, \ell_n = v'_n\}}{p}{B}$ or
  $\nomatch{\{\ell_1 = v_1, \ldots, \ell_n = v_n\}}{p}$ and
  $\nomatch{\{\ell_1 = v'_1, \ldots, \ell_n = v'_n\}}{p}$,
  by \eeqrule{matching}, we may conclude \[[\vec{v}/\vec{x}](
    \{\ell_1 = e_1, \ldots, \ell_n = e_n\}
    \eeq
    \{ \ell_1 = e'_1, \ldots, \ell_n = e'_n\}
  )\]\\\\
\noindent \isorule{application var/e}:
  Let $\Gamma = \vec{z} : \vec{\tau}$
  and let $\vec{v}$ be arbitrary where $v_i : \tau_i$ and $\val{v_i}$
  for all $v_i \in \vec{v}$.
  Assume $[\vec{v}/\vec{z}]
    \left(\forallval_{\Gamma'}.\sigma \right)
  $. It must be shown that $[\vec{v}/\vec{z}](
    x\ e_1 \eeq e_2
  )$.
  
  By the inductive hypothesis we have
  \[\forallval_{\Gamma, y : \tau} . \left(
    \text{if $\left(\forallval_{\Gamma'} . \sigma \right)$ then
    $y \eeq [y/(x\ e_1)]e_2 : \tau$}
  \right)\]
  
  Since we are only concerned with proving our approach sound over valuable expressions, without loss of generality, we can assume that $\evaluatesto{x\ e_1}{w}$ for some value $w$ such that $\Gamma \vdash w : \tau$ and $\val{w}$. Since $y$ is fresh, the inductive hypothesis written above implies
  
  \[
    \text{if $[w/y][\vec{v}/\vec{z}]\left(\forallval_{\Gamma'} . \sigma \right)$ then
    $[w/y][\vec{v}/\vec{z}](y \eeq [y/(x\ e_1)]e_2 : \tau)$}
  \]
  
  By assumption, we already have $[w/y][\vec{v}/\vec{z}]\left(\forallval_{\Gamma'} . \sigma \right)$. Therefore we have \[[w/y][\vec{v}/\vec{z}](y \eeq [y/(x\ e_1)]e_2 : \tau)\]
  
  which is equivalent to \[[\vec{v}/\vec{z}](w \eeq [w/(x\ e_1)]e_2 : \tau)\]
  
  Since $\evaluatesto{x\ e_1}{w}$, the two are extensionally equivalent. By the referential transparency of LambdaPix, the above expression is equivalent to
  \[[\vec{v}/\vec{z}](x\ e_1 \eeq [(x\ e_1)/(x\ e_1)]e_2 : \tau)\]
  
  which is simply \[[\vec{v}/\vec{z}](x\ e_1 \eeq e_2 : \tau)\]
  
\section{Experimental Results}\label{section:results}

We implemented our approach in a tool called \toolname to serve as a grading assistant by clustering equivalent programs into equivalence classes. The goal of our evaluation is to answer the following:

\begin{itemize}
    \item[Q1.] Can \toolname automatically identify equivalent programs in programming assignments for introductory functional programming courses? 
    \item[Q2.] How many equivalence classes are found by \toolname?
    \item[Q3.] What is the runtime performance of \toolname?
\end{itemize}

\subsection{Implementation}

\toolname is implemented in Standard ML and is publicly available as open-source at \url{https://github.com/CMU-TOP/zeus}. \toolname takes as input a set of homework assignments from an introductory functional programming course at the college level taught in Standard ML. Each submission is transpiled from Standard ML into LambdaPix, and then \toolname is run pairwise on the transpiled expressions and outputs a logical formula. If this formula is valid, then both expressions are algorithmically similar and guaranteed to be equivalent, so they are placed into the same equivalence class.
As an optimization, since extensional equivalence is transitive, if \toolname verifies that two programs $p_1$ and $p_2$ are (not) equivalent, and that $p_1$ is also (not) equivalent to $p_3$, then \toolname does not check that $p_2$ is equivalent to $p_3$. This optimization significantly reduces the number of comparisons that otherwise would be quadratic in the number of assignments.

In the definition of LambdaPix, we assumed an arbitrary fixed set of disjoint algebraic datatypes with unique associated injection labels. This is unrealistic for an implementation since Standard ML includes datatype declarations. Our transpilation from Standard ML to LambdaPix instead scrapes all datatype declarations from the original Standard ML submission and uses those datatypes and their constructors as LambdaPix's set of datatypes and injection labels.

To determine the validity of the formulas generated by \toolname, we use the SMT solver Z3~\cite{moura-z3-tacas08} using the theory of quantifier-free linear integer arithmetic and the theory of datatypes. From the theory of quantifier-free linear integer arithmetic, we use the built-in functions ``$+$'', ``$-$'', ``$*$'', ``$\leq$'', ``$<$'', ``$\geq$'', and ``$>$'', corresponding to the primitive operations of LambdaPix. We use the theory of datatypes to represent base terms of all types aside from \emph{ints} and \emph{booleans}. 

Although we only use these two theories in \toolname, nothing restricts a different implementation from using additional theories. For instance, another implementation could leverage the theory of \emph{strings} by adding strings as a base type in LambdaPix and adding the SMT solver's built-in string functions to LambdaPix's set of primitive operations.

\subsection{Benchmarks}

\begin{table}[!t]
\begin{center}
\caption{Description of homework assignments used in our evaluation}
\vspace*{-1mm}
\label{tbl:benchmarks}
\begin{tabular}{|l|p{3.5cm}|p{7.0cm}|}
\hline
Function & Signature & Description\\
\hline
\sf concat & \sf int list list $\rightarrow$ int list & 
\textsf{concat} takes a list of \textsf{int lists} and returns their concatenation without using the built-in ``@'' function\\
\hline
\sf prefixSum & \sf int list $\rightarrow$ int list & \textsf{prefixSum} replaces each \emph{i-th} element in an \textsf{int list} with the sum of the list's first $i+1$ elements\\
\hline
\sf countNonZero & \sf int tree $\rightarrow$ int & \textsf{countNonZero} takes an \textsf{int tree T} and returns the number of nonzero nodes in \textsf{T}\\
\hline
\sf quicksort & \sf ('a * 'a $\rightarrow$ order) * 'a list $\rightarrow$ 'a list & \textsf{quicksort} implements the quicksort algorithm \\
\hline
\sf slowDoop & \sf ('a * 'a $\rightarrow$ order) * 'a list $\rightarrow$ 'a list & \textsf{slowDoop} takes a comparison function and uses it to remove all duplicates in a list\\
\hline
\sf differentiate & \sf (int $\rightarrow$ real) $\rightarrow$ (int $\rightarrow$ real) & 
\textsf{differentiate} differentiates a polynomial that is represented with the type \textsf{int} $\rightarrow$ \textsf{real}\\
\hline
\sf integrate & \sf (int $\rightarrow$ real) $\rightarrow$ real $\rightarrow$ (int $\rightarrow$ real) & \textsf{integrate} takes a polynomial \textsf{p} and a real \textsf{c} and returns the antiderivative of \textsf{p} with constant of integration \textsf{c}\\
\hline
\sf treefoldr & \sf ('a * 'b $\rightarrow$ 'b) $\rightarrow$ 'b $\rightarrow$ 'a tree $\rightarrow$ 'b & (\textsf{treefoldr} \textsf{g init T}) returns (foldr \textsf{g init L}) where \textsf{L} is the inorder traversal of \textsf{T}\\
\hline
\sf treeReduce & \sf ('a * 'a $\rightarrow$ 'a) $\rightarrow$ 'a $\rightarrow$ 'a tree $\rightarrow$ 'a & \textsf{treeReduce} is the same as \textsf{treefoldr} except that it must have $O(log~n)$ span assuming \textsf{g} is associative and \textsf{init} is an identity for \textsf{g}\\
\hline
\sf findN & ('a $\rightarrow$ bool) $\rightarrow$ ('a * 'a $\rightarrow$ bool) $\rightarrow$ 'a shrub $\rightarrow$ int $\rightarrow$ ('a list $\rightarrow$ 'b) $\rightarrow$ (unit $\rightarrow$ 'b) $\rightarrow$ 'b & (\textsf{findN p eq T n s k}) returns \textsf{s [x1, …, xn]} where \textsf{[x1, …, xn]} are the leftmost values for \textsf{T} such that for all $i$ from $1$ to $n$, \textsf{p xi} returns true and the \textsf{xi}'s are eq-distinct. (\textsf{findN p eq T n s k}) returns \textsf{k()} if no such \textsf{[x1, …, xn]} exist\\
\hline
\sf sat & prop $\rightarrow$ ((string * bool) list $\rightarrow$ 'a) $\rightarrow$ (unit $\rightarrow$ 'a) $\rightarrow$ 'a & \textsf{sat} takes in a proposition, a success function \textsf{s} from a list assigning booleans to free variables to \textsf{'a}, and a failure function from \textsf{unit} to \textsf{'a}. If the proposition is satisfiable by an assignment of free variables \textsf{A}, then \textsf{sat} returns \textsf{s(A)}. Otherwise, it returns \textsf{k()}\\
\hline
\sf findPartition & 'a list $\rightarrow$ ('a list $\rightarrow$ bool) $\rightarrow$ ('a list $\rightarrow$ bool) $\rightarrow$ bool & (\textsf{findParition} A pL pR) returns true if there exist an \textsf{L} and \textsf{R} such that \textsf{(L, R)} is a partition of \textsf{A} where \textsf{pL} accepts \textsf{L} and \textsf{pR} accepts \textsf{R}. (\textsf{findPartition A pL pR}) returns false otherwise\\
\hline
\end{tabular}
\end{center}
\vspace*{-4mm}
\end{table} 

To evaluate \toolname, we used more than 4,000 student submissions from an introductory functional programming course. The number of submissions varies between 318 and 351 per assignment. Table~\ref{tbl:benchmarks} describes the twelve assignments that were used in our evaluation. These assignments show a large diversity of programs that includes different datatypes and the use of pattern matching and are a good test suite to test the applicability of \toolname as a grading assistant. Figure~\ref{fig:datatypes} shows some of the datatype declarations that are assumed by the homework assignments presented in Table~\ref{tbl:benchmarks}.

\begin{figure}[h]
\begin{lstlisting}
datatype 'a tree = Empty | Node of 'a tree * 'a * 'a tree
datatype 'a shrub = Leaf of 'a | Branch of 'a shrub * 'a shrub
datatype prop = Const of bool | Var of string | Not of prop
              | And of prop * prop | Or of prop * prop
              
\end{lstlisting}
\caption{Datatype declarations assumed by homework assignments}
\label{fig:datatypes}
\end{figure}

\subsection{Clustering of equivalent programs}

\begin{table}[!t]
\begin{center}
\caption{Analysis of the number of equivalent classes (ECs)}
\vspace{-2mm}
\label{tbl:results}
\begin{tabular}{|l|c|c|c|c|c|c|}
\hline
& \# & ECs & \multicolumn{1}{p{1.5cm}|}{\thead{\normalsize 90th \\\normalsize Percentile\\ \normalsize ECs}} & \multicolumn{1}{p{1.5cm}|}{\thead{\normalsize 75th \\\normalsize Percentile\\ \normalsize ECs}} & \multicolumn{1}{p{1.5cm}|}{\thead{\normalsize Non-\\ \normalsize singleton\\ \normalsize ECs}} & \multicolumn{1}{p{1.5cm}|}{\thead{\normalsize \% in Non-\\ \normalsize singleton \\ \normalsize ECs}}\\ \hline
\sf treefoldr & 332 & 22 & 3 & 1 & 9 & 96 \\ \hline
\sf integrate & 323 & 34 & 4 & 1 & 5 & 91\\ \hline
\sf slowDoop & 347 & 30 & 6 & 2 & 12 & 95\\ \hline
\sf countNonZero & 351 & 29 & 8 & 4 & 13 & 95\\ \hline
\sf concat & 351 & 40 & 8 & 2 & 10 & 91\\ \hline
\sf treeReduce & 332 & 57 & 24 & 8 & 20 & 89\\ \hline
\sf prefixSum & 351 & 68 & 33 & 7 & 23 & 87\\ \hline
\sf differentiate & 316 & 65 & 34 & 3 & 6 & 81 \\ \hline
\sf quicksort & 347 & 73 & 39 & 9 & 18 & 84\\ \hline
\sf findN & 330 & 73 & 40 & 7 & 18 & 83 \\ \hline
\sf findPartition & 331 & 83 & 50 & 12 & 22 & 82\\ \hline
\sf sat & 318 & 104 & 73 & 25 & 14 & 72\\ \hline
\end{tabular}
\end{center}
\end{table}

\begin{table}[!t]
\begin{center}
\caption{Analysis of correctness of student submissions}
\vspace{-2mm}
\label{tbl:correctnessresults}
\begin{tabular}{|l|c|c|c|c|c|c|}
\hline
& \multicolumn{1}{p{1.8cm}|}{\thead{\normalsize Correct \\\normalsize Submissions}} & \multicolumn{1}{p{1.3cm}|}{\thead{\normalsize Correct \\\normalsize ECs}} & \multicolumn{1}{p{1.4cm}|}{\thead{\normalsize Non-\\ \normalsize singleton\\ \normalsize Correct\\ \normalsize ECs}} & \multicolumn{1}{p{1.8cm}|}{\thead{\normalsize Incorrect \\\normalsize Submissions}} & \multicolumn{1}{p{1.4cm}|}{\thead{\normalsize Incorrect \\\normalsize ECs}} & \multicolumn{1}{p{1.4cm}|}{\thead{\normalsize Non-\\ \normalsize singleton\\ \normalsize Incorrect\\ \normalsize ECs}}\\ \hline
\sf treefoldr & 302 & 12 & 4  & 30  & 10 & 5   \\ \hline
\sf integrate & 307 & 23 & 3  & 16  & 11 & 2   \\ \hline
\sf slowDoop & 346 & 29 & 12 & 1   & 1  & 0   \\ \hline
\sf countNonZero & 346 & 26 & 12 & 5   & 3  & 1 \\ \hline
\sf concat & 336 & 30 & 9  & 15  & 10 & 1   \\ \hline
\sf treeReduce & 188 & 18 & 8  & 144 & 39 & 12  \\ \hline
\sf prefixSum & 347 & 64 & 23 & 4   & 4  & 0   \\ \hline
\sf differentiate & 308 & 59 & 5  & 8   & 6  & 1 \\ \hline
\sf quicksort & 328 & 56 & 16 & 19  & 17 & 2   \\ \hline
\sf findN & 296 & 48 & 14 & 34  & 25 & 4   \\ \hline
\sf findPartition & 291 & 59 & 13 & 40  & 24 & 9  \\ \hline 
\sf sat & 273 & 72 & 11 & 45  & 32 & 3   \\ \hline
\end{tabular}
\end{center}
\end{table}

Tables ~\ref{tbl:results} and ~\ref{tbl:correctnessresults} analyze the equivalent classes detected by \toolname. In particular, for each task, Table ~\ref{tbl:results} shows the number of submissions (\#), the number of equivalent classes (ECs), the number of equivalence classes that contain 90\% and 75\% of the submissions (90th and 75th Percentile ECs, respectively), the number of equivalent classes containing more than 1 submission (Non-singleton ECs), and the percentage of submissions found equivalent to at least one other submissions (\% in Non-singleton ECs). Table ~\ref{tbl:correctnessresults} shows the number of correct and incorrect student submissions, the number of equivalence classes containing only correct or incorrect submissions, and the number of equivalence classes containing multiple correct or incorrect submissions. There were no equivalence classes that contained both correct and incorrect submissions.

A common trend among all tasks was that a significant majority of student submissions were placed into a relatively small number of large equivalence classes, with the remaining submissions widely dispersed among many small equivalence classes, frequently of size 1. For instance, for the task \textsf{concat}, \toolname detected 40 equivalent classes. However, only 10 of those classes contain more than one submission, and 8 equivalence classes contain more than 90\% of the submissions. In almost all tasks, the largest equivalence classes consisted of various distinct but correct solutions to the problem. The one exception to this trend was that in the task treeReduce, a significant number of students mistook associativity for commutativity or otherwise assumed that the function passed into treeReduce was necessarily commutative. This common misunderstanding resulted in a large number of incorrect submissions for treeReduce, but because the misunderstanding was common, \toolname was still able to place the majority of incorrect submissions into a small number of large equivalence classes. In all tasks, at least 72\% of submissions were identified as equivalent to at least one other submission. These results support the hypothesis that \toolname can be used as a grading assistant to reduce the workload of instructors in reviewing equivalent code, thus freeing their time to provide more detailed feedback.

\subsection{Runtime performance}

Table~\ref{tbl:time} shows the time needed by \toolname to cluster all assignments for a given task when running on a common Mac laptop with a 1.6GHz processor and 4 GB of RAM. 
Specifically, for each task, it shows the number of submissions, the total time to cluster submissions in seconds, the number of pairwise comparisons performed during clustering, and the average time for a single comparison in seconds.
The average time to compare two individual submissions is small and it ranges from 0.046 seconds to 0.074 seconds. When performing the clustering of a given assignment, we can observe that the number of comparisons is much less than quadratic and that the total time varies between 1 and 8 minutes. This shows that \toolname is efficient in practice and can be used in real-time to help instructors grade assignments.

\begin{table}[!t]
\begin{center}
\caption{Runtime analysis}
\vspace{-1mm}
\label{tbl:time}
\centering
\begin{tabular}{|l|c|c|c|c|}
\hline
 & \# & Total Time (s)& \multicolumn{1}{p{2.6cm}|}{\thead{\normalsize Number of\\ \normalsize Comparisons}} & Average Time (s) \\ \hline
\sf treefoldr & 332 & 33.701  & 694  & 0.049  \\ \hline
\sf integrate & 323 & 45.833  & 991  & 0.046  \\ \hline
\sf slowDoop & 347 & 57.564  & 1,201 & 0.048  \\ \hline
\sf countNonZero & 351 & 72.268  & 1,578 & 0.046  \\ \hline
\sf concat & 351 & 76.110  & 1,614 & 0.047  \\ \hline
\sf treeReduce & 332 & 146.830 & 3,089 & 0.048  \\ \hline
\sf prefixSum & 351 & 205.695 & 4,112 & 0.050  \\ \hline
\sf differentiate & 316 & 140.797 & 3,025 & 0.047  \\ \hline
\sf quicksort & 347 & 210.554 & 4,303 & 0.049  \\ \hline
\sf findN & 330 & 202.577 & 3,660 & 0.055  \\ \hline
\sf findPartition & 331 & 284.502 & 4,807 & 0.059 \\ \hline
\sf sat & 318 & 486.218 & 6,532 & 0.074  \\ \hline
\end{tabular}
\end{center}
\end{table}

\subsection{Discussion}

We manually inspected the cases where \toolname did not put two programs in the same equivalence class. The most common reasons for this were the following:

\begin{itemize}
    \item \emph{The two programs are not equivalent:} since these programs correspond to actual student submissions, not all of the programs are correct. When an incorrect implementation produces the wrong output on any number of inputs, our algorithm appropriately puts it in a different equivalence class from the correct submissions.
    Additionally, for the \textsf{sat} task, the correct behavior of this function when an input proposition is satisfiable by multiple assignments is not fully defined. If multiple assignments \textsf{A} satisfy the proposition, there are no rules about which \textsf{A} to use when returning \textsf{s(A)}. So for this task, two correct submissions could produce different outputs.
    \item \emph{The two programs use different recursive helper functions:} we found cases where equivalence classes were distinguished by the structure of the helper functions students created. Since our current inference rules do not consider these cases, \toolname fails to recognize that two programs are equivalent if they use recursive helper functions with different input structures.
    \item \emph{The two recursive programs use different base cases:} our algorithm's treatment of fixed points causes it to never peer into a recursive call. Our algorithm's treatment of case expressions causes it to only recognize two expressions as equivalent if they handle all inputs in basically the same way. Together, these have the implication that when one expression treats a certain input as a base case while the other expression treats it as a recursive case, then the algorithm will be unable to recognize the expressions as equivalent.
    \item \emph{One of the programs uses built-in Standard ML functions:} seven out of the twelve tasks involve list manipulation operations. For instance, the top five tasks with the largest number of equivalence classes (\textsf{sat, findPartition, findN, quicksort,} and \textsf{prefixSum}) correspond to tasks that involve list manipulation. Many of the submissions for these tasks use built-in Standard ML functions for list reversal or list concatenation. We did not use a theory of list structures in our SMT Solver, so we were only able to recognize two expressions as equivalent if they used these built-in functions on the same input inputs and order or if they did not use these built-in functions at all.
\end{itemize}

We note that even with the current limitations, \toolname already shows that it can efficiently cluster the majority of the submissions into a few equivalence classes. Also, \toolname could be extended by adding additional inference rules or support for additional SMT theories that would allow the identification of equivalent programs that are currently missed by \toolname.

\section{Related Work}
\label{sec:related}

Proving that two problems are equivalent is a well-studied topic and has many applications ranging from hardware equivalence~\cite{blt-functional-1989}, compiler optimizations~\cite{zpf+translation-2002}, to program equivalence~\cite{godlin-regression-verification-dac09}. However, the use of program equivalence for grading programming assignments is scarce~\cite{kaleeswaran-coderassist-fse16}. In this section, we cover related work from program equivalence and automatic grading that is closer to our approach.

\subsection{Program Equivalence}

\paragraph{Program Verification.} The problem of program equivalence can be reduced to a verification problem by showing that both programs satisfy the same specification. For instance, model-checking techniques~\cite{DBLP:journals/fmsd/ClarkeBRZ01,ckl-ansic-2004} can be used to show that two C programs satisfy the same specification. This specification can be written to ensure that for the same input, the programs are equivalent if they always produce the same output. 
Fedyukovich et al.~\cite{fgs-equivalence-2016} present techniques for proving that two similar programs have the same property rather than being equivalent. Their approach requires formally verifying one of the programs and using this proof to check the validity of the property in the other program by establishing a coupling between the two programs. 
A similar approach can also be done for functional programs. For instance, one could write a formal specification of the functionality of a program in Why3ML~\cite{DBLP:journals/sttt/BobotFMP15}. We tried this approach by writing a formal specification for programming assignments for the function \texttt{concat}, however, the Why3 framework~\cite{DBLP:journals/sttt/BobotFMP15} was not able to prove that the program satisfied the specification. In general, proving the program equivalence concerning a specification is a more challenging task than the one we address in this paper since we can take advantage of program structure to prove that they are equivalent.

\paragraph{Regression Verification.} In regression verification~\cite{godlin-regression-verification-dac09,felsign-verification-ase14}, the goal is to prove that two versions of a program are equivalent. One approach is to transform loops in programs to recursive procedures and to match the recursive calls in both programs and abstract them via uninterpreted functions~\cite{godlin-regression-verification-dac09}. Other approaches use invariant inference techniques to prove the equivalence of programs with loops~\cite{felsign-verification-ase14}. By using these techniques, one can encode the two versions of the program into Horn clauses and use constraint solvers to automatically find certain kinds of invariants. Alternatively, one can also use symbolic execution and static analysis to generate summaries of program behaviors that capture the modifications between the programs. These summaries can be encoded into logical formulas and their equivalence can be checked using SMT solvers~\cite{backes-summaries-spin13}.
Our approaches also consider that student submissions are similar but they are not different versions of the same program. Even though we do not use any invariant generation techniques, this is orthogonal to our approach and could increase the number of equivalent classes detected for recursive programs.

\paragraph{Contextual Equivalence} 

There is a broad set of work that targets contextual equivalence for functional programs. Approaches based on step-indexed logical relations~\cite{dreyer-step-equivalence-lics09,ahmed-state-context-popl09,ahmed-esop06} or on bisimulations~\cite{pierce-bisimulation-popl05,koutavas-bisimulation-popl06,hur-bisimulation-popl12} have been used to prove context equivalence of functional programs with different fragments of ML that often include finite datatypes and integer references. 
While these approaches are more theoretical and focus on functional programs with state, we do not support state but can handle pattern matching which is crucial for a practical tool to cluster programming assignments of introductory functional courses.
The closest approach to ours is the one recently presented by Jaber~\cite{jaber-syteci-2020}. Jaber presents techniques for checking the equivalence of OCaml programs with state. His approach focuses in particular on contextual equivalence and developing a framework in which references can be properly accounted for. Our approach neglects references, as we require programs to be purely functional, but includes a more comprehensive treatment of datatypes. We attempted to compare our \toolname's performance against Jaber's {\sc SyTeCi}\xspace prototype, but unfortunately, all of our benchmarks included datatypes that were not supported by the available prototype.

\subsection{Automatic grading}

\paragraph{Clustering similar assignments.} 
To help instructors to grade programming assignments, several automatic techniques have been proposed to cluster similar assignments into buckets with the purpose of giving automatic feedback~\cite{gulwani-clustering-pldi18,wang-sarfgen-pldi18,pu-ml-splash16,kaleeswaran-coderassist-fse16}. Our approach differs from these since our goal is not to replace the instructor or to fully automate the grading but rather to use \toolname as a grading assistant with formal guarantees. 

\textsf{CLARA}~\cite{gulwani-clustering-pldi18} cluster correct programs and selects a canonical program from each cluster to be considered as the reference solution. In this approach, a pair of programs $p_1$ and $p_2$ are said to be dynamic equivalent if they have the same control-flow and if related variables in $p_1$ and $p_2$ always have the same values, in the same order, during the program execution on the same inputs. In contrast, our approach has a stronger notion of equivalence since we do not depend on dynamic program analysis. 
\textsf{CodeAssist}~\cite{kaleeswaran-coderassist-fse16} clusters submissions for dynamic programming assignments by their solution strategy. They consider a small set of features and if two programs share these features then they are put in the same cluster.
Other clustering approaches are based on deep learning techniques~\cite{pu-ml-splash16} and also provide no formal guarantees about the quality of the clustering.
\textsf{SemCluster}~\cite{perry-semcluster-pldi19} improves upon other clustering techniques by considering semantic program features. They use control flow features and data flow features to represent each program and merge this information to create a program feature vector. K-means clustering is used to cluster all programs based on the program feature vectors. Even though there are no formal guarantees for the equivalence of programs in each cluster, experimental results~\cite{perry-semcluster-pldi19} show that the number of clusters found by \textsf{SemCluster} is much smaller than competitive approaches.

\paragraph{Automatic repair.} 
\textsf{AutoGrader}~\cite{singh-autograder-pldi13} takes as input a reference solution and an error model that consists of potential corrections and uses constraint solving techniques to find a minimum number of corrections that can be used to repair the incorrect student solution.
\textsf{Sarfgen}~\cite{wang-sarfgen-pldi18} uses a three-stage algorithm based on search, align, and repair. It starts by searching for a small number of correct programs that can be used to repair the incorrect submission and have the same control-flow structure. Next, they compute a syntactic distance between those programs using an embedding of ASTs into numerical vectors. These programs are then aligned and the differences between aligned statements can suggest corrections that can be repaired automatically.

Automatic repair is better suited for Massive Open Online Courses where a fully automated method is needed, while our approach is better suited for large, in-person courses, where the feedback of instructors can be more beneficial. The feedback returned by automatic repair tools is limited to changes in the code, while our approach is meant to assist instructors to provide more detailed feedback for students. Each equivalent class will have specific comments that are more helpful to the student than a repaired version of their submission. Moreover, while our approach can be used for both correct and incorrect submissions, automatic repair is only useful to fix incorrect submissions and cannot give any feedback for different implementations of correct submissions.

\paragraph{Formal guarantees.} Liu et al.~\cite{liu-grading-formal-icse19} proposes to automatically determine the correctness of an assignment against a reference solution. Instead of using test cases, they use symbolic execution to search for semantically different execution paths between a student's submission and the reference solution. If such paths exist, then the submission is considered incorrect and feedback can be provided by using counterexamples based on path deviations. 
Our approach is not based on symbolic execution but instead uses inference rules to derive a formula for which both student submissions are equivalent if and only if they have the same structure and the observable behavior.

\textsf{CodeAssist}~\cite{kaleeswaran-coderassist-fse16} checks equivalence of a candidate submission from a cluster with a correct solution of that cluster that has been previously validated by an instructor. They exploit the fact of just handling dynamic programming assignments to establish a correspondence between variables and control locations of the two programs. Using this correspondence, they can encode the problem into SMT and prove program equivalence. Our approach is more general since our inference rules simulate relationships between expressions of the two programs and can be applied to several problem domains and not just dynamic programming assignments.

\section{Conclusion}

We present techniques for checking for equivalence between purely functional programs. Guided by inference rules that inform needed equivalences between two programs' subexpressions, our approach simultaneously deconstructs the expressions being compared to build up a formula that is valid only if the expressions are equivalent. We prove the soundness of our approach: if our approach takes in two expressions of the same type and outputs a valid formula, then the two expressions are equivalent. We implement our approach and show that it can assist grading by clustering over 4,000 real student code submissions from an introductory functional programming class taught at the undergraduate level.

\appendix
\section{Appendix}
\label{sec:appendix}

Here we include the full proof that our approach is sound. This proof makes use of a few lemmas:\\
\newcounter{lemma}
\newcommand{\definelemma}[1]{\refstepcounter{lemma}\label{lemma:#1}Lemma \thelemma}
\newcommand{\Lemma}[1]{Lemma \ref{lemma:#1}}
\textbf{\definelemma{contextsdisjoint}}:
  if $\Gamma \vdash e_1 \smtequiv{\sigma} e_2 : \tau \dashv \Gamma'$
  or $\Gamma \vdash e_1 \smtequivnormal{\sigma} e_2 : \tau \dashv \Gamma'$,
  then $\Gamma$ and $\Gamma'$ are disjoint.
  \textit{Proof}: by induction on
  $\Gamma \vdash e_1 \smtequiv{\sigma} e_2 : \tau \dashv \Gamma'$ and
  $\Gamma \vdash e_1 \smtequivnormal{\sigma} e_2 : \tau \dashv \Gamma'$\\
\textbf{\definelemma{whnfequiv}}:
  WHNF reduction preserves equivalence:\\
  if
  $e_1' \eeq e_2'$,
  $e_1 \whnfbig e_1'$, and
  $e_2 \whnfbig e_2'$, then
  $e_1 \eeq e_2$.
  \textit{Proof}:
  By induction on $e \whnfstep e'$ and appealing to the dynamics,
  we have that if
  $e_1' \eeq e_2'$,
  $e_1 \whnfstep e_1'$, and
  $e_1 \whnfstep e_2'$ then
  $e_1 \eeq e_2$.
  The rest goes through by induction on $e \whnfbig e'$.\\
\textbf{\definelemma{equiv}}:
  If $e_1 \equiv e_2$, $e_1 \term$, $e_2 \term$, and $e_1$ and $e_2$ are both closed expressions (as opposed to patterns or open terms), then $e_1 \eeq e_2$.
  \textit{Proof}: By induction on $e \term$ and from the definition of $\equiv$ given in section 4.1.\\
\textbf{\definelemma{eeqmatch}}:
  If $e_1 \eeq e_2 : \tau$,
  $\evaluatesto{e_1}{v_1}$, and
  $\evaluatesto{e_2}{v_2}$,
  then for all patterns $p$ where
  $p :: \tau \dashv \Gamma$,
  it is the case that
  either $\matches{v_1}{p}{B}$ and $\matches{v_2}{p}{B}$ or
  $\nomatch{v_1}{p}$ and $\nomatch{v_2}{p}$.
  \textit{Proof}: by induction on $e_1 \eeq e_2 : \tau$.
  If $\tau = \tau_1 \rightarrow \tau_2$ then by inversion of
  $p :: \tau \dashv \Gamma$, $p$ must either be a wildcard or a variable.
  Then by \matchrule{wildcard} and \matchrule{variable}, we have that
  $\matches{v_1}{p}{B}$ and $\matches{v_2}{p}{B}$.
  If $\tau$ is not an arrow type, then we conclude by $\eeqrule{matching}$.\\
\textbf{\definelemma{whnfatomicval}}:
  If $\Gamma \vdash e : \tau$, $e \term$, and $e$ is in weak head normal form,
  then $\forallval_\Gamma . (\val{e}$ or $e = o\ e'$) for some primitive operation $o$ and expression $e'$.
  \textit{Proof}: by induction on $e \term$ and appealing to the dynamics.\\
\textbf{\definelemma{freshmatch}}:
  If
  $\Gamma \vdash v : \tau$,
  $p :: \tau \dashv \Gamma'$,
  $\Gamma, \Gamma' \vdash e : \tau'$,
  $\matches{v}{p}{B}$,
  and
  $\freshen{p}{e}{p'}{e'}$,
  then
  $\matches{v}{p'}{B'}$ and $[B]e = [B']e'$.
  \textit{Proof}: by induction on $\freshen{p}{e}{p'}{e'}$.\\
\textbf{\definelemma{matcheq}}:
  If $\matches{v}{p}{B}$ then $v \equiv [B]p$.
  \textit{Proof}: by induction on $\matches{v}{p}{B}$ and from the definition of $\equiv$.
\textbf{\definelemma{matchneq}}:
  If $\nomatch{v}{p}$ then $v \not\equiv p$.
  \textit{Proof}: by induction on $\nomatch{v}{p}$ and from the definition of $\equiv$.\\
\textbf{\definelemma{equatebindings1}}:
  If $\Gamma \vdash v : \tau$, $p_1 :: \tau \dashv \Gamma'$, $\Gamma, \Gamma' \vdash e : \tau'$, and $\equatebindings{p_1.e_1}{p_2.e_2}{p'.e'_1}{p'.e'_2}$ then ($\matches{v}{p_1}{B}$, $\matches{v}{p'}{B'}$ and $[B]e = [B']e'$) or ($\nomatch{v}{p_1}$ and $\nomatch{v}{p'}$). \textit{Proof}: by induction on $\equatebindings{p_1.e_1}{p_2.e_2}{p'.e'_1}{p'.e'_2}$ and appealing to $\matches{v}{p}{B}$ and $\nomatch{v}{p}$.\\
\textbf{\definelemma{equatebindings2}}:
  If $\Gamma \vdash v : \tau$, $p_2 :: \tau \dashv \Gamma'$, $\Gamma, \Gamma' \vdash e : \tau'$, and $\equatebindings{p_1.e_1}{p_2.e_2}{p'.e'_1}{p'.e'_2}$ then ($\matches{v}{p_2}{B}$, $\matches{v}{p'}{B'}$ and $[B]e = [B']e'$) or ($\nomatch{v}{p_2}$ and $\nomatch{v}{p'}$). \textit{Proof}: by induction on $\equatebindings{p_1.e_1}{p_2.e_2}{p'.e'_1}{p'.e'_2}$ and appealing to $\matches{v}{p}{B}$ and $\nomatch{v}{p}$.\\
\textbf{\definelemma{freshentogethermatch1}}:
   If $\Gamma \vdash v : \tau$ and $\freshentogether{\{p_1.e_1 \mid \ldots \mid p_n.e_n\}}{\{\ldots\}}{\{p'_1.e'_1 \mid \ldots \mid p'_m.e'_n\}}{\{\ldots'\}}{s}$ then for all $i \in [n]$, either $(\matches{v}{p_i}{B}$, $\matches{v}{p_i'}{B'}$ and $[B][e_i] = B'[e_i'])$ or $(\nomatch{v}{p_i}$ and $\nomatch{v}{p_i'}$). \textit{Proof}: by induction on $\mathsf{FT}$ and appealing to \Lemma{equatebindings1}.\\
\textbf{\definelemma{freshentogethermatch2}}:
   If $\Gamma \vdash v : \tau$ and $\freshentogether{\{\ldots\}}{\{p_1.e_1 \mid \ldots \mid p_m.e_m\}}{\{\ldots'\}}{\{p'_1.e'_1 \mid \ldots \mid p'_m.e'_m\}}{s}$ then for all $j \in [m]$, either $(\matches{v}{p_j}{B}$, $\matches{v}{p_j'}{B'}$ and $[B][e_j] = B'[e_j'])$ or $(\nomatch{v}{p_j}$ and $\nomatch{v}{p_j'}$). \textit{Proof}: by induction on $\mathsf{FT}$ and appealing to \Lemma{equatebindings2}.\\
\textbf{\definelemma{freshentogether}}:
  If $\Gamma \vdash v : \tau$ and $\freshentogether{\{M\}}{\{M'\}}{\{p_1.e_1 \mid \ldots \mid p_n.e_n\}}{\{p'_1.e'_1 \mid \ldots \mid p'_m.e'_m\}}{s}$ then $\texttt{case}\ v\ \{M\} \eeq \texttt{case}\ v\ \{p_1.e_1 \mid \ldots \mid p_n.e_n\}$ and $\texttt{case}\ v\ \{M'\} \eeq \texttt{case}\ v\ \{p'_1.e'_1 \mid \ldots \mid p'_n.e'_n\}$. \textit{Proof}: by Lemmas \ref{lemma:freshentogethermatch1} and \ref{lemma:freshentogethermatch2} and appealing to \dynrule{case2}.\\
\textbf{\definelemma{freshentogether s}}:
  If $\freshentogether{\{M\}}{\{M'\}}{\{p_1.e_1 \mid \ldots \mid p_n.e_n\}}{\{p'_1.e'_1 \mid \ldots \mid p'_m.e'_m\}}{s}$, then for all $i \leq s$, $p_i = p_i'$. \textit{Proof}: By induction on $\mathsf{FT}$ and the fact that $\equatebindings{p_1.e_1}{p_2.e_2}{p'.e_1'}{p'.e_2'}$ uses the same $p'$ in both of the new bindings.\\

\noindent Using these lemmas, we proceed to prove the soundness of each rule. We use the following inductive hypothesis:
\begin{itemize}
\item
  If $\Gamma \vdash e_1 \smtequiv{\sigma} e_2 : \tau \dashv \Gamma'$ then
  $\forallval_{\Gamma} . \left(
    \text{if $\left(\forallval_{\Gamma'} . \sigma \right)$ then $e_1 \eeq e_2 : \tau$}
  \right)$.
\item
  If $\Gamma \vdash e_1 \smtequivnormal{\sigma} e_2 : \tau \dashv \Gamma'$ then
  $\forallval_{\Gamma} . \left(
    \text{if $\left(\forallval_{\Gamma'} . \sigma \right)$ then $e_1 \eeq e_2 : \tau$}
  \right)$.
\end{itemize}
In cases where the exact type of the expressions are obvious or irrelevant,
we use the shorthand $e \eeq e'$ to mean that $e \eeq e' : \tau$ for some type $\tau$.\\

\begin{itemize}
\item \theisoexprule:
  Let $\Gamma = \vec{x} : \vec{\tau}$
  and let $\vec{v}$ be arbitrary where $v_i : \tau_i$ and $\val{v_i}$
  for all $v_i \in \vec{v}$.
  Assume $[\vec{v}/\vec{x}]
    \left(\forallval_{\Gamma'}.\sigma\right)
  $. It must be shown that $[\vec{v}/\vec{x}](e_1 \eeq e_2)$.

  By the inductive hypothesis, we have that
  $\forallval_{\Gamma} . \left(
    \text{if $\left(\forallval_{\Gamma'} . \sigma \right)$ then $e'_1 \eeq e'_2$}
  \right)$,
  and therefore
  \[[\vec{v}/\vec{x}]\left(
    \text{if $\left(\forallval_{\Gamma'} . \sigma \right)$ then $e'_1 \eeq e'_2$}
  \right)\]
  equivalently,
  \[
    \text{if $[\vec{v}/\vec{x}]\left(\forallval_{\Gamma'} . \sigma \right)$ then $[\vec{v}/\vec{x}](e'_1 \eeq e'_2)$}
  \]
  As we have $[\vec{v}/\vec{x}]\left(\forallval_{\Gamma'} . \sigma \right)$ by assumption,
  we may conclude $[\vec{v}/\vec{x}](e'_1 \eeq e'_2)$. By \Lemma{whnfequiv}, this implies that $[\vec{v}/\vec{x}](e_1 \eeq e_2)$ as desired.
\item \isorule{atomic}:
  Let $\Gamma = \vec{x} : \vec{\tau}$
  and let $\vec{v}$ be arbitrary where $v_i : \tau_i$ and $\val{v_i}$
  for all $v_i \in \vec{v}$.
  Assume $[\vec{v}/\vec{x}]
    \left(\forallval_{\cdot}.e_1 \equiv e_2\right)
  $, or equivalently $[\vec{v}/\vec{x}](e_1 \equiv e_2)$. It must be shown that $[\vec{v}/\vec{x}](e_1 \eeq e_2)$.

  As $[\vec{v}/\vec{x}](e_1 \equiv e_2)$, $e_1 \term$, and $e_2 \term$, we have $[\vec{v}/\vec{x}](e_1 \eeq e_2)$ by \Lemma{equiv}.
\item \isorule{record}:
  Let $\Gamma = \vec{x} : \vec{\tau}$
  and let $\vec{v}$ be arbitrary where $v_i : \tau_i$ and $\val{v_i}$
  for all $v_i \in \vec{v}$.
  Assume $[\vec{v}/\vec{x}]
    \left(\forallval_{\Gamma'_1, \ldots, \Gamma'_n}.\sigma_1 \land \ldots \sigma_n\right)
  $. It must be shown that $[\vec{v}/\vec{x}](
    \{\ell_1 = e_1, \ldots, \ell_n = e_n\}
    \eeq
    \{ \ell_1 = e'_1, \ldots, \ell_n = e'_n\}
  )$.

  By conjunction and that all the $\Gamma'_i$ are disjoint, we have that
  for all $i \in [n]$, $\forallval_{\Gamma'_i} . \sigma_i$.
  Then by the inductive hypotheses, we have that $[\vec{v}/\vec{x}](e_i \eeq e_i' : \tau_i)$.
  Since we are only concerned with proving our approach sound over valuable expressions, without loss of generality, we can assume that $\evaluatesto{[\vec{v}/\vec{x}]e_i}{v_i}$ and $\evaluatesto{[\vec{v}/\vec{x}]e'_i}{v'_i}$ for some values $v_i$ and $v_i'$.
  By \Lemma{eeqmatch} we have that for all $p_i$ where $p_i :: \tau_i \dashv \Gamma_i$,
  either $\matches{v_i}{p_i}{B_i}$ and $\matches{v'_i}{p_i}{B_i}$ or $\nomatch{v_i}{p_i}$ and 
  $\nomatch{v'_i}{p_i}$.\\

  To appeal to \eeqrule{matching},
  let $p$ be an arbitrary pattern such that
  $p :: \{\ell_1 : \tau_1, \ldots, \ell_n : \tau_n\} \dashv \Gamma'$.
  We proceed by cases:
  \begin{itemize}
  \item In the case that for all $i \in [n]$
    $\matches{v_i}{p_i}{B_i}$ and $\matches{v'_i}{p_i}{B_i}$,
    by \matchrule{record1} we have
    $\matches{\{\ell_1 = v_1, \ldots, \ell_n = v_n\}}{p}{B_1 \ldots B_n}$ and
    $\matches{\{\ell_1 = v'_1, \ldots, \ell_n = v'_n\}}{p}{B_1 \ldots B_n}$.
  \item In the case that there is some $i \in [n]$ where
    $\nomatch{v_i}{p_i}$ and $\nomatch{v'_i}{p_i}$,
    by \matchrule{record2} we have
    $\nomatch{\{\ell_1 = v_1, \ldots, \ell_n = v_n\}}{p}$ and
    $\nomatch{\{\ell_1 = v'_1, \ldots, \ell_n = v'_n\}}{p}$.
  \end{itemize}
  Since in all cases either
  $\matches{\{\ell_1 = v_1, \ldots, \ell_n = v_n\}}{p}{B}$ and
  $\matches{\{\ell_1 = v'_1, \ldots, \ell_n = v'_n\}}{p}{B}$ or
  $\nomatch{\{\ell_1 = v_1, \ldots, \ell_n = v_n\}}{p}$ and
  $\nomatch{\{\ell_1 = v'_1, \ldots, \ell_n = v'_n\}}{p}$,
  by \eeqrule{matching}, we may conclude \[[\vec{v}/\vec{x}](
    \{\ell_1 = e_1, \ldots, \ell_n = e_n\}
    \eeq
    \{ \ell_1 = e'_1, \ldots, \ell_n = e'_n\}
  )\]
\item \isorule{projection}:
  Let $\Gamma = \vec{x} : \vec{\tau}$
  and let $\vec{v}$ be arbitrary where $v_i : \tau_i$ and $\val{v_i}$
  for all $v_i \in \vec{v}$.
  Assume $[\vec{v}/\vec{x}]
    \left(\forallval_{\Gamma'}.\sigma\right)
  $. It must be shown that $[\vec{v}/\vec{x}](
    e_1 \cdot \ell_i
    \eeq
    e_2 \cdot \ell_i
  )$.

  By the inductive hypothesis, we have
  $\forallval_{\Gamma} . \left(
    \text{if $\left(\forallval_{\Gamma'} . \sigma \right)$ then $e_1 \eeq e_2$}
  \right)$,
  and therefore
  \[[\vec{v}/\vec{x}]\left(
    \text{if $\left(\forallval_{\Gamma'} . \sigma \right)$ then $e_1 \eeq e_2$}
  \right)\]
  equivalently,
  \[
    \text{if $[\vec{v}/\vec{x}]\left(\forallval_{\Gamma'} . \sigma \right)$ then $[\vec{v}/\vec{x}](e_1 \eeq e_2)$}
  \]
  As we have $[\vec{v}/\vec{x}]\left(\forallval_{\Gamma'} . \sigma \right)$ by assumption,
  we may conclude \[
    [\vec{v}/\vec{x}](e_1 \eeq e_2)
  \]

  As extensional equivalence is the same as contextual equivalence,
  from this we may conclude \[
    [\vec{v}/\vec{x}](e_1 \cdot \ell_i \eeq e_2 \cdot \ell_i)
  \]
\item \isorule{injection}:
  For the same reasons as in the proof for \isorule{projection},
  we have that $[\vec{v}/\vec{x}](e_1 \eeq e_2)$.
  As extensional equivalence is the same as contextual equivalence, we may then conclude \[
    [\vec{v}/\vec{x}](i \cdot e_1 \eeq i \cdot e_2)
  \]
\item \isorule{lambda}:
  Let $\Gamma = \vec{y} : \vec{\tau}$
  and let $\vec{v}$ be arbitrary where $v_i : \tau_i$ and $\val{v_i}$
  for all $v_i \in \vec{v}$.
  Assume $[\vec{v}/\vec{y}]
    \left(\forallval_{x:\tau,\Gamma'}.\sigma \right)
  $. It must be shown that $[\vec{v}/\vec{y}](
    \lambda x_1 . e_1
    \eeq
    \lambda x_2 . e_2
  )$.

  To appeal to \eeqrule{extensionality},
  take arbitrary $w$ such that $\Gamma \vdash w : \tau$ and $\val{w}$.
  By the inductive hypothesis, we have
  \[\forallval_{\Gamma, x:\tau} . \left(
    \text{if $\left(\forallval_{\Gamma'} . \sigma \right)$ then
    $[x/x_1]e_1 \eeq [x/x_2]e_2 : \tau'$}
  \right)\]
  and therefore, since $x$ is fresh,
  \[
    \text{if $[w/x][\vec{v}/\vec{y}]\left(\forallval_{\Gamma'} . \sigma \right)$ then
    $[w/x][\vec{v}/\vec{y}]([x/x_1]e_1 \eeq [x/x_2]e_2 : \tau')$}
  \]
  By assumption, we already have $[w/x][\vec{v}/\vec{y}]\left(\forallval_{\Gamma'} . \sigma \right)$.
  Therefore we have \[[w/x][\vec{v}/\vec{y}]([x/x_1]e_1 \eeq [x/x_2]e_2 : \tau')\]

  By \dynrule{beta}, we have \[
    \stepsto{[\vec{v}/\vec{y}](\lambda x_1 . e_1)\ w}{[\vec{v}/\vec{y}][w/x_1]e_1 = [\vec{v}/\vec{y}][w/x][x/x_1]e_1}\]
  and
  \[\stepsto{[\vec{v}/\vec{y}](\lambda x_2 . e_2)\ w}{[\vec{v}/\vec{y}][w/x_2]e_2 = [\vec{v}/\vec{y}][w/x][x/x_2]e_2}\]
  Since $[w/x][\vec{v}/\vec{y}]([x/x_1]e_1 \eeq [x/x_2]e_2 : \tau')$ and $x$ is fresh,
  we have that $[\vec{v}/\vec{y}](
    \lambda x_1 . e_1
    \eeq
    \lambda x_2 . e_2
  )$ by \eeqrule{extensionality}.
\item \isorule{fix}:
  Let $\Gamma = \vec{y} : \vec {\tau}$ and let $\vec{v}$ be arbitrary where $v_i : \tau_i$ and $\val{v_i}$ for all $v_i \in \vec{v}$. Assume $[\vec{v}/\vec{y}]
    \left(\forallval_{x:\tau,\Gamma'}.\sigma \right)
  $. It must be shown that $[\vec{v}/\vec{y}](\texttt{fix}\ x_1\ \texttt{is}\ e_1 \eeq \texttt{fix}\ x_2\ \texttt{is}\ e_2)$. 
  
  Note that as long as $x$ is fresh, $[\vec{v}/\vec{y}](\texttt{fix}\ x_1\ \texttt{is}\ e_1 \eeq \texttt{fix}\ x\ \texttt{is}\ [x/x_1]e_1)$ and $[\vec{v}/\vec{y}](\texttt{fix}\ x_2\ \texttt{is}\ e_2 \eeq \texttt{fix}\ x\ \texttt{is}\ [x/x_2]e_2)$ by alpha equivalence. Since we have as a premise for \isorule{fix} that $x \fresh$, to show $[\vec{v}/\vec{y}](\texttt{fix}\ x_1\ \texttt{is}\ e_1 \eeq \texttt{fix}\ x_2\ \texttt{is}\ e_2)$, it suffices to show $[\vec{v}/\vec{y}](\texttt{fix}\ x\ \texttt{is}\ [x/x_1]e_1 \eeq \texttt{fix}\ x\ \texttt{is}\ [x/x_2]e_2)$. By the inductive hypothesis, we have
  \[\forallval_{\Gamma, x:\tau} . \left(
    \text{if $\left(\forallval_{\Gamma'} . \sigma \right)$ then
    $[x/x_1]e_1 \eeq [x/x_2]e_2 : \tau'$}
  \right)\]
  This implies that:
  \[\forallval_{x:\tau} . [\vec{v}/\vec{y}]\left(
    \text{if $\left(\forallval_{\Gamma'} . \sigma \right)$ then
    $[x/x_1]e_1 \eeq [x/x_2]e_2 : \tau'$}
  \right)\]
  which can be rearranged to:
  \[\text{if $[\vec{v}/\vec{y}]\left(\forallval_{x: \tau, \Gamma'} . \sigma \right)$ then
    $[\vec{v}/\vec{y}]\left(\forallval_{x:\tau}([x/x_1]e_1 \eeq [x/x_2]e_2 : \tau')\right)$}
  \]
  By assumption, we have that $[\vec{v}/\vec{y}]\left(\forallval_{x:\tau, \Gamma'} . \sigma \right)$, so this implies
  \[[\vec{v}/\vec{y}]\left(\forallval_{x:\tau}([x/x_1]e_1 \eeq [x/x_2]e_2 : \tau')\right)\]
  Since LambdaPix enjoys referential transparency, we can therefore substitute $[x/x_1]e_1$ for $[x/x_2]e_2$ in $[\vec{v}/\vec{y}]\texttt{fix}\ x\ \texttt{is}\ [x/x_1]e_1$ to get $[\vec{v}/\vec{y}](\texttt{fix}\ x\ \texttt{is}\ [x/x_1]e_1 \eeq \texttt{fix}\ x\ \texttt{is}\ [x/x_2]e_2)$. As $[\vec{v}/\vec{y}](\texttt{fix}\ x\ \texttt{is}\ [x/x_1]e_1 \eeq \texttt{fix}\ x_1\ \texttt{is}\ e_1)$ and $[\vec{v}/\vec{y}](\texttt{fix}\ x\ \texttt{is}\ [x/x_2]e_2 \eeq \texttt{fix}\ x_2\ \texttt{is}\ e_2)$, this implies $[\vec{v}/\vec{y}](\texttt{fix}\ x_1\ \texttt{is}\ e_1 \eeq \texttt{fix}\ x_2\ \texttt{is}\ e_2)$ as desired.
\item \isorule{application}:
  Let $\Gamma = \vec{x} : \vec{\tau}$
  and let $\vec{v}$ be arbitrary where $v_i : \tau_i$ and $\val{v_i}$
  for all $v_i \in \vec{v}$.
  Assume $[\vec{v}/\vec{x}]
    \left(\forallval_{\Gamma', \Gamma''}.\sigma \land \sigma'\right)
  $. It must be shown that $[\vec{v}/\vec{x}](
    e_1\ e_1'
    \eeq
    e_2\ e_2'
  )$.
  
  By the inductive hypothesis we have
  \[\forallval_{\Gamma} . \left(
    \text{if $\left(\forallval_{\Gamma'} . \sigma \right)$ then
    $e_1 \eeq e_2 : \tau \rightarrow \tau'$}
  \right)\]
  and therefore
  \[\text{if $[\vec{v}/\vec{x}]\left(\forallval_{\Gamma'} . \sigma \right)$ then
    $[\vec{v}/\vec{x}](e_1 \eeq e_2 : \tau \rightarrow \tau')$}
  \]
  As $\sigma$ does not contain any variables in $\Gamma''$,
  by assumption and conjunction we already have
  $[\vec{v}/\vec{x}]\left(\forallval_{\Gamma'} . \sigma \right)$
  therefore we may conclude
  \[[\vec{v}/\vec{x}](e_1 \eeq e_2 : \tau \rightarrow \tau')\]

  Similarly,
  by the inductive hypothesis we have
  \[\forallval_{\Gamma} . \left(
    \text{if $\left(\forallval_{\Gamma''} . \sigma' \right)$ then
    $e_1' \eeq e_2' : \tau$}
  \right)\]
  and therefore
  \[\text{if $[\vec{v}/\vec{x}]\left(\forallval_{\Gamma''} . \sigma' \right)$ then
    $[\vec{v}/\vec{x}](e_1' \eeq e_2' : \tau)$}
  \]
  As $\sigma'$ does not contain any variables in $\Gamma'$,
  by assumption and conjunction we already have
  $[\vec{v}/\vec{x}]\left(\forallval_{\Gamma''} . \sigma' \right)$
  therefore we may conclude
  \[[\vec{v}/\vec{x}](e_1' \eeq e_2' : \tau)\]

  Since $[\vec{v}/\vec{x}]e_1 \eeq [\vec{v}/\vec{x}]e_2 : \tau \rightarrow \tau'$, 
  by \eeqrule{extensionality} we have
  \[[\vec{v}/\vec{x}](e_1\ e_1') \eeq [\vec{v}/\vec{x}](e_2\ e_1') : \tau'\]
  Since $[\vec{v}/\vec{x}](e_1' \eeq e_2' : \tau)$,
  by referential transparency we then have
  \[[\vec{v}/\vec{x}](e_1\ e_1') \eeq [\vec{v}/\vec{x}](e_2\ e_2') : \tau'\]
\item \isorule{application var/e}:
  Let $\Gamma = \vec{z} : \vec{\tau}$
  and let $\vec{v}$ be arbitrary where $v_i : \tau_i$ and $\val{v_i}$
  for all $v_i \in \vec{v}$.
  Assume $[\vec{v}/\vec{z}]
    \left(\forallval_{\Gamma'}.\sigma \right)
  $. It must be shown that $[\vec{v}/\vec{z}](
    x\ e_1 \eeq e_2
  )$.
  
  By the inductive hypothesis we have
  \[\forallval_{\Gamma, y : \tau} . \left(
    \text{if $\left(\forallval_{\Gamma'} . \sigma \right)$ then
    $y \eeq [y/(x\ e_1)]e_2 : \tau$}
  \right)\]
  
  Since we are only concerned with proving our approach sound over valuable expressions, without loss of generality, we can assume that $\evaluatesto{x\ e_1}{w}$ for some value $w$ such that $\Gamma \vdash w : \tau$ and $\val{w}$. Since $y$ is fresh, the inductive hypothesis written above implies
  
  \[
    \text{if $[w/y][\vec{v}/\vec{z}]\left(\forallval_{\Gamma'} . \sigma \right)$ then
    $[w/y][\vec{v}/\vec{z}](y \eeq [y/(x\ e_1)]e_2 : \tau)$}
  \]
  
  By assumption, we already have $[w/y][\vec{v}/\vec{z}]\left(\forallval_{\Gamma'} . \sigma \right)$. Therefore we have \[[w/y][\vec{v}/\vec{z}](y \eeq [y/(x\ e_1)]e_2 : \tau)\]
  
  which is equivalent to \[[\vec{v}/\vec{z}](w \eeq [w/(x\ e_1)]e_2 : \tau)\]
  
  Since $\evaluatesto{x\ e_1}{w}$, the two are extensionally equivalent. By the referential transparency of LambdaPix, the above expression is equivalent to
  \[[\vec{v}/\vec{z}](x\ e_1 \eeq [(x\ e_1)/(x\ e_1)]e_2 : \tau)\]
  
  which is simply \[[\vec{v}/\vec{z}](x\ e_1 \eeq e_2 : \tau)\]
\item \isorule{application e/var}: By symmetry and \isorule{application var/e}.
\item \isorule{application f/e}: The proof for this is the same as in \isorule{application var/e} with $x$ substituted for $o$ (the fact that $x$ is a variable is never used in the proof of \isorule{application var/e}).
\item \isorule{application e/f}: By symmetry and \isorule{application f/e}.
\item \isorule{application fix/e}: The proof for this is the same as in \isorule{application var/e} with $x$ substituted for $\texttt{fix}\ x_1\ \texttt{is}\ e_1$, $e_1$ substituted for $e_2$, and $e_2$ substituted for $e$ (the fact that $x$ is a variable is never used in the proof of \isorule{application var/e}).
\item \isorule{application e/fix}: By symmetry and \isorule{application fix/e}.
\item \isorule{casel}:
  Let $\Gamma = \vec{x} : \vec{\tau}$
  and let $\vec{v}$ be arbitrary where $v_i : \tau_i$ and $\val{v_i}$
  for all $v_i \in \vec{v}$.
  Assume $[\vec{v}/\vec{x}]
    \forallval_{\Gamma_1, \Gamma'_1, \ldots, \Gamma_n, \Gamma'_n} .
    \land_{i \in [n]} ((\land_{j \in [i-1]} (e \not\equiv p_j')) \land e \equiv p_i') \Rightarrow \sigma_i
  $.
  
  It must be shown that $
    [\vec{v}/\vec{x}](\texttt{case}\ e\ \{ p_1 . e_1 \mid \ldots \mid p_n . e_n \} \eeq e' : \tau)
  $.

  By inversion of the statics we have that $\Gamma \vdash e : \tau'$. By assumption we have that $e \term$ and $e$ is in weak head normal form.
  Therefore by \Lemma{whnfatomicval} we have that either $\val{[\vec{v}/\vec{x}]e}$ or $[\vec{v}/\vec{x}]e = o\ e'$ for some primitive operation $o$ and expression $e'$. In the former case, let $w = [\vec{v}/\vec{x}]e$. Since we are only concerned with proving our approach sound over valuable expressions, in the latter case we can state without loss of generality that $\evaluatesto{o\ e'}{w}$ for some value $w$.

  Since case expressions are enforced to be exhaustive,
  there must be some $p_i$ such that $\matches{w}{p_i}{B}$ and for all $j < i$, $\nomatch{w}{p_j}$. By \Lemma{freshmatch}
  let $\matches{w}{p_i'}{B'}$.
  
  By our assumption and the semantics of conjunction we have that 
  \[
  [\vec{v}/\vec{x}]
    \forallval_{\Gamma_1, \Gamma'_1, \ldots, \Gamma_n, \Gamma'_n} .
    ((\land_{j \in [i-1]} (e \not\equiv p_j')) \land e \equiv p_i') \Rightarrow \sigma_i
  \]
  Since $\Gamma$ is disjoint with all other listed contexts, this is equivalent to:
  \[
    \forallval_{\Gamma_1, \Gamma'_1, \ldots, \Gamma_n, \Gamma'_n} .
    ((\land_{j \in [i-1]} [\vec{v}/\vec{x}](e \not\equiv p_j')) \land [\vec{v}/\vec{x}](e \equiv p_i')) \Rightarrow [\vec{v}/\vec{x}]\sigma_i
  \]
  Since $p_i'$ and each $p_j'$ only contains variables in $\Gamma_i$ or $\Gamma_j$ (and therefore none of $\vec{x}$), we have:
  \[
    \forallval_{\Gamma_1, \Gamma'_1, \ldots, \Gamma_n, \Gamma'_n} .
    ((\land_{j \in [i-1]} [\vec{v}/\vec{x}]e \not\equiv p_j') \land [\vec{v}/\vec{x}]e \equiv p_i') \Rightarrow [\vec{v}/\vec{x}]\sigma_i
  \]
  From our definition of $w$, this is equivalent to:
  \[
    \forallval_{\Gamma_1, \Gamma'_1, \ldots, \Gamma_n, \Gamma'_n} .
    ((\land_{j \in [i-1]} w \not\equiv p_j') \land w \equiv p_i') \Rightarrow [\vec{v}/\vec{x}]\sigma_i
  \]
  We may partially invoke this result with $B'$ which gives us: 
  \[
    [B']\forallval_{\Gamma_1, \Gamma'_1, \ldots, \Gamma_n, \Gamma'_n} .
    ((\land_{j \in [i-1]} w \not\equiv p_j') \land w \equiv p_i') \Rightarrow [\vec{v}/\vec{x}]\sigma_i
  \]
  Since the list of contexts is pairwise disjoint and since neither $w$ nor $p_j'$ for any $j < i$ contain any variables in $\Gamma_i$, we can rearrange this to get: 
  \[
    \forallval_{\Gamma_1, \Gamma'_1, \ldots, \Gamma_n, \Gamma'_n} .
    ((\land_{j \in [i-1]} w \not\equiv p_j') \land w \equiv [B']p_i') \Rightarrow [B'][\vec{v}/\vec{x}]\sigma_i
  \]
  By \Lemma{matcheq}, $w \equiv [B']p_i'$ is true, and by $i-1$ applications of \Lemma{matchneq}, $\land_{j \in [i-1]} w \not\equiv p_j'$ is true, so this is equivalent to:
  \[
  \forallval_{\Gamma_1, \Gamma'_1, \ldots, \Gamma_n, \Gamma'_n} . [B'][\vec{v}/\vec{x}]\sigma_i
  \]
  $\sigma_i$ only contains variables from $\Gamma$, $\Gamma_i$, and $\Gamma_i'$. Therefore, $[\vec{v}/\vec{x}]\sigma_i$ only contains variables from $\Gamma_i$, and $\Gamma_i'$. Therefore, $[B'][\vec{v}/\vec{x}]\sigma_i$ only contains variables from $\Gamma_i'$. Because of this, we can simplify and rearrange the above to:
  \[
  [B'][\vec{v}/\vec{x}] \forallval_{\Gamma_i'} . \sigma_i
  \]
  This allows us to invoke the inductive hypothesis to get $[B'][\vec{v}/\vec{x}]e'_i \eeq [B'][\vec{v}/\vec{x}]e' : \tau$.
  
  Since the variables in $B'$ don't appear in $e'$, this is equivalent to: $[B'][\vec{v}/\vec{x}]e'_i \eeq [\vec{v}/\vec{x}]e' : \tau$.
  
  By \Lemma{freshmatch}, this is equivalent to: $[B][\vec{v}/\vec{x}]e_i \eeq [\vec{v}/\vec{x}]e' : \tau$.

  By $\dynrule{case2}$, $\stepsto{
    [\vec{v}/\vec{x}]\left(\texttt{case}\ e\ \{ p_1 . e_1 \mid \ldots \mid p_n . e_n \}\right)
  }{
    [B][\vec{v}/\vec{x}]e_i
  }$.
  
  Therefore, since extensional equivalence is closed under evaluation:
  
  \[
    [\vec{v}/\vec{x}] (\texttt{case}\ e\ \{ p_1 . e_1 \mid \ldots \mid p_n . e_n \}
    \eeq
    e'
    : \tau)
  \]
\item $\isorule{caser}$:
  By symmetry and $\isorule{casel}$.
\item $\isorule{symcase}$
  Let $\Gamma = \vec{y} : \vec{\tau}$
  and let $\vec{v}$ be arbitrary where $v_i : \tau_i$ and $\val{v_i}$
  for all $v_i \in \vec{v}$.
  Assume $[\vec{v}/\vec{y}]
    \left(\forallval_{\Gamma', x:\tau', \Gamma''}.\sigma \land \sigma'\right)
  $. It must be shown that $[\vec{v}/\vec{y}](
    \texttt{case}\ e\ \{ \ldots \}
    \eeq
    \texttt{case}\ e'\ \{ \ldots \}
  )$.

  By the inductive hypothesis we have
  \[\forallval_{\Gamma} . \left(
    \text{if $\left(\forallval_{\Gamma'} . \sigma \right)$ then $e \eeq e'$}
  \right)\]
  and therefore
  \[\left(
    \text{if $[\vec{v}/\vec{y}]\left(\forallval_{\Gamma'} . \sigma \right)$ then $[\vec{v}/\vec{y}](e \eeq e')$}
  \right)\]
  Since $[\vec{v}/\vec{y}]\sigma$ only contains variables in $\Gamma'$,
  by assumption and conjunction we already have
  $[\vec{v}/\vec{y}]\left(\forallval_{\Gamma'} . \sigma \right)$.
  Therefore we may conclude $[\vec{v}/\vec{y}](e \eeq e')$.

  Since we are only concerned with proving our approach sound over valuable expressions, without loss of generality, we can assume that $\evaluatesto{[\vec{v}/\vec{y}]e}{w}$ for some value $w$ such that $\Gamma \vdash w : \tau'$ and $\val{w}$.
  By the inductive hypothesis we have
  \[\forallval_{\Gamma, x:\tau'} . \left(
    \text{if $\left(\forallval_{\Gamma''} . \sigma' \right)$ then
    $\texttt{case}\ x\ \{ \ldots \} \eeq \texttt{case}\ x\ \{ \ldots' \}$}
  \right)\]
  and therefore, since $x$ is fresh,
  \[\forallval_{x:\tau'} . \left(
    \text{if $[\vec{v}/\vec{y}]\left(\forallval_{\Gamma''} . \sigma' \right)$ then
    $[\vec{v}/\vec{y}](\texttt{case}\ x\ \{ \ldots \} \eeq \texttt{case}\ x\ \{ \ldots' \})$}
  \right)\]
  Invoking this with $w$, we have
  \[
    \text{if $[w/x][\vec{v}/\vec{y}]\left(\forallval_{\Gamma''} . \sigma' \right)$ then
    $[\vec{v}/\vec{y}](\texttt{case}\ w\ \{ \ldots \} \eeq \texttt{case}\ w\ \{ \ldots' \})$}
  \]
  By assumption and conjunction we already have
  $[w/x][\vec{v}/\vec{y}]\left(\forallval_{\Gamma''} . \sigma' \right)$.
  Therefore we may conclude \[[\vec{v}/\vec{y}](
    \texttt{case}\ w\ \{ \ldots \} \eeq \texttt{case}\ w\ \{ \ldots' \}
  )\]

  Since $[\vec{v}/\vec{y}]e \eeq w$ and $[\vec{v}/\vec{y}]e \eeq [\vec{v}/\vec{y}]e'$, we have $[\vec{v}/\vec{y}]e' \eeq w$ by transitivity. Then, by referential transparency and the above equivalence, we have:
  \[[\vec{v}/\vec{y}](
    \texttt{case}\ e\ \{ \ldots \} \eeq \texttt{case}\ e'\ \{ \ldots' \}
  )\]
\item \isorule{freshentogethercase1}:
  Let $\Gamma = \vec{y} : \vec{\tau}$
  and let $\vec{v}$ be arbitrary where $v_i : \tau_i$ and $\val{v_i}$
  for all $v_i \in \vec{v}$.
  Assume $[\vec{v}/\vec{y}] \forallval_{\Gamma_1, \Gamma_1' \ldots \Gamma_n, \Gamma_n'}. \Psi$. It must be shown that $
    [\vec{v}/\vec{y}](\texttt{case}\ x\ \{M\} \eeq \texttt{case}\ x\ \{M'\} : \tau)
  $.
  
  Since extensional equivalence is defined only over closed expressions, without loss of generality, we can assume that $[\vec{v}/\vec{y}]$ includes some binding for $x$ $[v/x]$. By partial application, $
    [\vec{v}/\vec{y}](\texttt{case}\ x\ \{M\} \eeq \texttt{case}\ x\ \{M'\} : \tau)
  $ is equivalent to $
    [\vec{v}/\vec{y}](\texttt{case}\ v\ \{M\} \eeq \texttt{case}\ v\ \{M'\} : \tau)
  $.
  
  By \Lemma{freshentogether}, $[\vec{v}/\vec{y}](\texttt{case}\ v\ \{M\} \eeq \texttt{case}\ v\ \{p_1.e_1 \mid \ldots \mid p_n.e_n\})$ and $[\vec{v}/\vec{y}](\texttt{case}\ v\ \{M'\} \eeq \texttt{case}\ v\ \{p'_1.e'_1 \mid \ldots \mid p'_m.e'_m\})$. So to show $
    [\vec{v}/\vec{y}](\texttt{case}\ v\ \{M\} \eeq \texttt{case}\ v\ \{M'\} : \tau)
  $, it suffices to show $
    [\vec{v}/\vec{y}](\texttt{case}\ v\ \{p_1.e_1 \mid \ldots \mid p_n.e_n\} \eeq \texttt{case}\ v\ \{p'_1.e'_1 \mid \ldots \mid p'_m.e'_m\} : \tau)
  $.
  
  Since case expressions are enforced to be exhaustive, there must be some $p_i$ and $p_j'$ such that $\matches{v}{p_i}{B}$ and for all $j < i$, $\nomatch{v}{p_j}$. Since $i \leq s$ or $i > s$, we must consider both cases.\\
  
  We first consider the case where $i \leq s$. 
  
  By our assumption that $[\vec{v}/\vec{y}]\Psi$ is valid and the semantics of conjunction, we have that 
  \[[\vec{v}/\vec{y}]\forallval_{\Gamma_1, \Gamma_1' \ldots \Gamma_n, \Gamma_n'}.(\land_{i \in [s]}\sigma_i)\] 
  so in particular, at the $i$ such that $\matches{v}{p_i}{B}$, we have\\ 
  \[[\vec{v}/\vec{y}]\forallval_{\Gamma_1, \Gamma_1' \ldots\Gamma_n, \Gamma_n'}. \sigma_i\] 

  Since $\sigma_i$ only contains variables from $\Gamma, \Gamma_i,$ and $\Gamma_i'$, this is equivalent to $[\vec{v}/\vec{y}]\forallval_{\Gamma_i, \Gamma_i'}.\sigma_i$. Then, by our inductive hypothesis, $[\vec{v}/\vec{y}]\forallval_{\Gamma_i}.(e_i \eeq e_i')$. This implies $[\vec{v}/\vec{y}][B](e_i \eeq e_i')$. Since $i \leq s$, by \Lemma{freshentogether s}, $\matches{v}{p_i}{B}$, $\matches{v}{p_i'}{B}$, and for all $j < i$, $\nomatch{v}{p_j}$ and $\nomatch{v}{p_j'}$. So by \dynrule{case2}, $\stepsto{[\vec{v}/\vec{y}](\texttt{case}\ v\ \{p_1.e_1 \mid \ldots \mid p_n.e_n\})} {[\vec{v}/\vec{y}][B]e_i}$ and $\stepsto{[\vec{v}/\vec{y}](\texttt{case}\ v\ \{p_1'.e_1' \mid \ldots \mid p_m'.e_m'\})} {[\vec{v}/\vec{y}][B]e_i'}$. Since extensional equivalence is closed under evaluation, this implies $
    [\vec{v}/\vec{y}](\texttt{case}\ v\ \{p_1.e_1 \mid \ldots \mid p_n.e_n\} \eeq \texttt{case}\ v\ \{p'_1.e'_1 \mid \ldots \mid p'_m.e'_m\} : \tau)
  $ as desired.\\
  
  This leaves the case where $i > s$.
  By our assumption that $[\vec{v}/\vec{y}]\Psi$ is valid and the semantics of conjunction, we have that $[\vec{v}/\vec{y}]\forallval_{\Gamma_1, \Gamma_1' \ldots \Gamma_n, \Gamma_n'}.(\land_{j \in [s+1, n]} ((\land_{k \in [j-1]} (x \not\equiv p_k)) \land x \equiv p_j) \Rightarrow \sigma_j)$, so in particular, when $j=i$ such that $\matches{v}{p_i}{B}$, we have:
  \[ [\vec{v}/\vec{y}]\forallval_{\Gamma_1, \Gamma_1' \ldots \Gamma_n, \Gamma_n'}.((\land_{k \in [i-1]} (x \not\equiv p_k)) \land x \equiv p_i) \Rightarrow \sigma_i\]
  
  Since $\Gamma$ is disjoint with all other listed contexts, this is equivalent to:
  
  \[ \forallval_{\Gamma_1, \Gamma_1' \ldots \Gamma_n, \Gamma_n'}.((\land_{k \in [i-1]} [\vec{v}/\vec{y}](x \not\equiv p_k)) \land [\vec{v}/\vec{y}](x \equiv p_i)) \Rightarrow [\vec{v}/\vec{y}]\sigma_i\]
  
  Since $p_i$ and $p_k$ only contain variables in $\Gamma_i$ and $\Gamma_k$ (and therefore, none of $\vec{y}$), we have:
  
  \[ \forallval_{\Gamma_1, \Gamma_1' \ldots \Gamma_n, \Gamma_n'}.((\land_{k \in [i-1]} [\vec{v}/\vec{y}]x \not\equiv p_k) \land [\vec{v}/\vec{y}]x \equiv p_i) \Rightarrow [\vec{v}/\vec{y}]\sigma_i\]
  
  Recalling that $[v/x]$ is included in $[\vec{v}/\vec{y}]$, this is equivalent to:
  
  \[ \forallval_{\Gamma_1, \Gamma_1' \ldots \Gamma_n, \Gamma_n'}.((\land_{k \in [i-1]} v \not\equiv p_k) \land v \equiv p_i) \Rightarrow [\vec{v}/\vec{y}]\sigma_i\]
  
  We may partially invoke this result with $B$ which gives us:
  
  \[ [B]\forallval_{\Gamma_1, \Gamma_1' \ldots \Gamma_n, \Gamma_n'}.((\land_{k \in [i-1]} v \not\equiv p_k) \land v \equiv p_i) \Rightarrow [\vec{v}/\vec{y}]\sigma_i\]
  
  Since the list of contexts is pairwise disjoint and since neither $v$ nor $p_k$ for any $k<i$ contain any variables in $\Gamma_i$, we can rearrange this to get:
  
  \[ \forallval_{\Gamma_1, \Gamma_1' \ldots \Gamma_n, \Gamma_n'}.((\land_{k \in [i-1]} v \not\equiv p_k) \land v \equiv [B]p_i) \Rightarrow [B][\vec{v}/\vec{y}]\sigma_i\]
  
  By \Lemma{matcheq} $v \equiv [B]p_i$ is true and by $i-1$ applications of \Lemma{matchneq} $\land_{k \in [i-1]} v \not\equiv p_k$ is true, so this is equivalent to:
  
  \[ \forallval_{\Gamma_1, \Gamma_1' \ldots \Gamma_n, \Gamma_n'}. [B][\vec{v}/\vec{y}]\sigma_i\]
  
  $\sigma$ only contains variables from $\Gamma, \Gamma_i,$ and $\Gamma_i'$. Therefore, $[\vec{v}/\vec{y}]\sigma_i$ only contains variables from $\Gamma_i$ and $\Gamma_i'$. Therefore, $[B][\vec{v}/\vec{y}]\sigma_i$ only contains variables from $\Gamma_i'$. Because of this, we can simplify and rearrange the above to:
  
  \[ [B][\vec{v}/\vec{y}]\forallval_{\Gamma_i'}. \sigma_i\]
  
  This allows us to invoke the inductive hypothesis to get \[
  [B][\vec{v}/\vec{y}]e_i \eeq [B][\vec{v}/\vec{y}]\texttt{case}\ x\ \{p'_1.e'_1 \mid \ldots \mid p'_m.e'_m\} : \tau
  \]
  
  Since the variables in $B$ don't appear in $\texttt{case}\ x\ \{M'\}$, and since $[v/x]$ is included in $[\vec{v}/\vec{y}]$, this is equivalent to
  \[
  [B][\vec{v}/\vec{y}]e_i \eeq [\vec{v}/\vec{y}]\texttt{case}\ v\ \{p'_1.e'_1 \mid \ldots \mid p'_m.e'_m\} : \tau
  \]
  
  By \dynrule{case2}, $\stepsto{\texttt{case}\ v\ \{p_1.e_1 \mid \ldots \mid p_n.e_n\}}{[B][\vec{v}/\vec{y}]e_i}$.
  
  Therefore, since extensional equivalence is closed under evaluation:
  
  \[
    [\vec{v}/\vec{x}] (\texttt{case}\ v\ \{p_1.e_1 \mid \ldots \mid p_n.e_n\}
    \eeq
    \texttt{case}\ v\ \{p'_1.e'_1 \mid \ldots \mid p'_m.e'_m\}
    : \tau)
  \]
  
  So regardless of whether $i \leq s$ or $i > s$, we have $[\vec{v}/\vec{x}] (\texttt{case}\ v\ \{p_1.e_1 \mid \ldots \mid p_n.e_n\}
    \eeq
    \texttt{case}\ v\ \{p'_1.e'_1 \mid \ldots \mid p'_m.e'_m\}
    : \tau)$ as desired
\item \isorule{freshentogethercase2}: By symmetry and \isorule{freshentogethercase1}.\\
\end{itemize}
We have verified the soundness of each rule. Therefore, by induction: 
\begin{itemize}
\item
  If $\Gamma \vdash e_1 \smtequiv{\sigma} e_2 : \tau \dashv \Gamma'$ then
  $\forallval_{\Gamma} . \left(
    \text{if $\left(\forallval_{\Gamma'} . \sigma \right)$ then $e_1 \eeq e_2 : \tau$}
  \right)$.
\item
  If $\Gamma \vdash e_1 \smtequivnormal{\sigma} e_2 : \tau \dashv \Gamma'$ then
  $\forallval_{\Gamma} . \left(
    \text{if $\left(\forallval_{\Gamma'} . \sigma \right)$ then $e_1 \eeq e_2 : \tau$}
  \right)$.
\end{itemize}

The above statements imply the soundness theorem. When $\Gamma_{\textnormal{initial}} \vdash e_1 \smtequiv{\sigma} e_2 : \tau \dashv \Gamma'$
we have $\forallval_{\Gamma_{\textnormal{initial}}} . \left(\text{if $\left(\forallval_{\Gamma'} . \sigma \right)$ then $e_1 \eeq e_2 : \tau$}\right)$. Since $\Gamma_{\textnormal{initial}}$ contains only primitive operations, which are omitted from the $\forallval_\Gamma.j$ judgment, the outer quantifier quantifies over no variables, so we have that
$\text{if $\left(\forallval_{\Gamma'} . \sigma \right)$ then $e_1 \eeq e_2 : \tau$}$.
This together with the assumption that $\forallval_{\Gamma'} . \sigma$
allows us to conclude that $e_1 \eeq e_2 : \tau$. So for all expressions $e_1$ and $e_2$, if $\Gamma_{\textnormal{initial}} \vdash e_1 \smtequiv{\sigma} e_2 : \tau \dashv \Gamma'$ and
$\forallval_{\Gamma'} . \sigma$, then $e_1 \eeq e_2 : \tau$.

\section*{Acknowledgments}
This work was partially funded by National Science Foundation (Grants CCF-1901381, CCF-1762363, and CCF-1629444).

\bibliography{main.bib}


\begin{thebibliography}{24}


\ifx \showCODEN    \undefined \def \showCODEN     #1{\unskip}     \fi
\ifx \showDOI      \undefined \def \showDOI       #1{#1}\fi
\ifx \showISBNx    \undefined \def \showISBNx     #1{\unskip}     \fi
\ifx \showISBNxiii \undefined \def \showISBNxiii  #1{\unskip}     \fi
\ifx \showISSN     \undefined \def \showISSN      #1{\unskip}     \fi
\ifx \showLCCN     \undefined \def \showLCCN      #1{\unskip}     \fi
\ifx \shownote     \undefined \def \shownote      #1{#1}          \fi
\ifx \showarticletitle \undefined \def \showarticletitle #1{#1}   \fi
\ifx \showURL      \undefined \def \showURL       {\relax}        \fi
\providecommand\bibfield[2]{#2}
\providecommand\bibinfo[2]{#2}
\providecommand\natexlab[1]{#1}
\providecommand\showeprint[2][]{arXiv:#2}

\bibitem[\protect\citeauthoryear{Ahmed, Dreyer, and Rossberg}{Ahmed
  et~al\mbox{.}}{2009}]%
        {ahmed-state-context-popl09}
\bibfield{author}{\bibinfo{person}{Amal Ahmed}, \bibinfo{person}{Derek Dreyer},
  {and} \bibinfo{person}{Andreas Rossberg}.} \bibinfo{year}{2009}\natexlab{}.
\newblock \showarticletitle{State-dependent representation independence}. In
  \bibinfo{booktitle}{\emph{Proc. Symposium on Principles of Programming
  Languages}}. \bibinfo{publisher}{{ACM}}, \bibinfo{pages}{340--353}.
\newblock
\urldef\tempurl%
\url{https://doi.org/10.1145/1480881.1480925}
\showDOI{\tempurl}


\bibitem[\protect\citeauthoryear{Ahmed}{Ahmed}{2006}]%
        {ahmed-esop06}
\bibfield{author}{\bibinfo{person}{Amal~J. Ahmed}.}
  \bibinfo{year}{2006}\natexlab{}.
\newblock \showarticletitle{Step-Indexed Syntactic Logical Relations for
  Recursive and Quantified Types}. In \bibinfo{booktitle}{\emph{Proc. European
  Symposium on Programming}}. \bibinfo{publisher}{Springer},
  \bibinfo{pages}{69--83}.
\newblock
\urldef\tempurl%
\url{https://doi.org/10.1007/11693024\_6}
\showDOI{\tempurl}


\bibitem[\protect\citeauthoryear{Backes, Person, Rungta, and Tkachuk}{Backes
  et~al\mbox{.}}{2013}]%
        {backes-summaries-spin13}
\bibfield{author}{\bibinfo{person}{John~D. Backes}, \bibinfo{person}{Suzette
  Person}, \bibinfo{person}{Neha Rungta}, {and} \bibinfo{person}{Oksana
  Tkachuk}.} \bibinfo{year}{2013}\natexlab{}.
\newblock \showarticletitle{Regression Verification Using Impact Summaries}. In
  \bibinfo{booktitle}{\emph{Proc. International Symposium Model Checking
  Software}}. \bibinfo{publisher}{Springer}, \bibinfo{pages}{99--116}.
\newblock
\urldef\tempurl%
\url{https://doi.org/10.1007/978-3-642-39176-7\_7}
\showDOI{\tempurl}


\bibitem[\protect\citeauthoryear{Berman and Trevillyan}{Berman and
  Trevillyan}{1989}]%
        {blt-functional-1989}
\bibfield{author}{\bibinfo{person}{C~Leonard Berman} {and}
  \bibinfo{person}{Louise~H Trevillyan}.} \bibinfo{year}{1989}\natexlab{}.
\newblock \showarticletitle{Functional comparison of logic designs for VLSI
  circuits}. In \bibinfo{booktitle}{\emph{Proc. International Conference on
  Computer-Aided Design}}. IEEE, \bibinfo{pages}{456--459}.
\newblock
\urldef\tempurl%
\url{https://doi.org/10.1109/ICCAD.1989.76990}
\showDOI{\tempurl}


\bibitem[\protect\citeauthoryear{Bobot, Filli{\^{a}}tre, March{\'{e}}, and
  Paskevich}{Bobot et~al\mbox{.}}{2015}]%
        {DBLP:journals/sttt/BobotFMP15}
\bibfield{author}{\bibinfo{person}{Fran{\c{c}}ois Bobot},
  \bibinfo{person}{Jean{-}Christophe Filli{\^{a}}tre}, \bibinfo{person}{Claude
  March{\'{e}}}, {and} \bibinfo{person}{Andrei Paskevich}.}
  \bibinfo{year}{2015}\natexlab{}.
\newblock \showarticletitle{Let's verify this with Why3}.
\newblock \bibinfo{journal}{\emph{Int. J. Softw. Tools Technol. Transf.}}
  \bibinfo{volume}{17}, \bibinfo{number}{6} (\bibinfo{year}{2015}),
  \bibinfo{pages}{709--727}.
\newblock
\urldef\tempurl%
\url{https://doi.org/10.1007/s10009-014-0314-5}
\showDOI{\tempurl}


\bibitem[\protect\citeauthoryear{Clarke, Kroening, and Lerda}{Clarke
  et~al\mbox{.}}{2004}]%
        {ckl-ansic-2004}
\bibfield{author}{\bibinfo{person}{Edmund Clarke}, \bibinfo{person}{Daniel
  Kroening}, {and} \bibinfo{person}{Flavio Lerda}.}
  \bibinfo{year}{2004}\natexlab{}.
\newblock \showarticletitle{A tool for checking ANSI-C programs}. In
  \bibinfo{booktitle}{\emph{Proc. International Conference on Tools and
  Algorithms for the Construction and Analysis of Systems}}. Springer,
  \bibinfo{pages}{168--176}.
\newblock
\urldef\tempurl%
\url{https://doi.org/10.1007/978-3-540-24730-2\_15}
\showDOI{\tempurl}


\bibitem[\protect\citeauthoryear{Clarke, Biere, Raimi, and Zhu}{Clarke
  et~al\mbox{.}}{2001}]%
        {DBLP:journals/fmsd/ClarkeBRZ01}
\bibfield{author}{\bibinfo{person}{Edmund~M. Clarke}, \bibinfo{person}{Armin
  Biere}, \bibinfo{person}{Richard Raimi}, {and} \bibinfo{person}{Yunshan
  Zhu}.} \bibinfo{year}{2001}\natexlab{}.
\newblock \showarticletitle{Bounded Model Checking Using Satisfiability
  Solving}.
\newblock \bibinfo{journal}{\emph{Formal Methods Syst. Des.}}
  \bibinfo{volume}{19}, \bibinfo{number}{1} (\bibinfo{year}{2001}),
  \bibinfo{pages}{7--34}.
\newblock
\urldef\tempurl%
\url{https://doi.org/10.1023/A:1011276507260}
\showDOI{\tempurl}


\bibitem[\protect\citeauthoryear{de~Moura and Bj{\o}rner}{de~Moura and
  Bj{\o}rner}{2008}]%
        {moura-z3-tacas08}
\bibfield{author}{\bibinfo{person}{Leonardo~Mendon{\c{c}}a de Moura} {and}
  \bibinfo{person}{Nikolaj Bj{\o}rner}.} \bibinfo{year}{2008}\natexlab{}.
\newblock \showarticletitle{{Z3:} An Efficient {SMT} Solver}. In
  \bibinfo{booktitle}{\emph{Proc. International Conference on Tools and
  Algorithms for the Construction and Analysis of Systems}}.
  \bibinfo{publisher}{Springer}, \bibinfo{pages}{337--340}.
\newblock
\urldef\tempurl%
\url{https://doi.org/10.1007/978-3-540-78800-3\_24}
\showDOI{\tempurl}


\bibitem[\protect\citeauthoryear{Dreyer, Ahmed, and Birkedal}{Dreyer
  et~al\mbox{.}}{2009}]%
        {dreyer-step-equivalence-lics09}
\bibfield{author}{\bibinfo{person}{Derek Dreyer}, \bibinfo{person}{Amal Ahmed},
  {and} \bibinfo{person}{Lars Birkedal}.} \bibinfo{year}{2009}\natexlab{}.
\newblock \showarticletitle{Logical Step-Indexed Logical Relations}. In
  \bibinfo{booktitle}{\emph{Proc. Annual Symposium on Logic in Computer
  Science}}. \bibinfo{publisher}{{IEEE} Computer Society},
  \bibinfo{pages}{71--80}.
\newblock
\urldef\tempurl%
\url{https://doi.org/10.1109/LICS.2009.34}
\showDOI{\tempurl}


\bibitem[\protect\citeauthoryear{Fedyukovich, Gurfinkel, and
  Sharygina}{Fedyukovich et~al\mbox{.}}{2016}]%
        {fgs-equivalence-2016}
\bibfield{author}{\bibinfo{person}{Grigory Fedyukovich}, \bibinfo{person}{Arie
  Gurfinkel}, {and} \bibinfo{person}{Natasha Sharygina}.}
  \bibinfo{year}{2016}\natexlab{}.
\newblock \showarticletitle{{Property Directed Equivalence via Abstract
  Simulation}}. In \bibinfo{booktitle}{\emph{Proc. International Conference
  Computer-Aided Verification}}. \bibinfo{publisher}{Springer},
  \bibinfo{pages}{433--453}.
\newblock
\urldef\tempurl%
\url{https://doi.org/10.1007/978-3-319-41540-6\_24}
\showDOI{\tempurl}


\bibitem[\protect\citeauthoryear{Felsing, Grebing, Klebanov, R{\"{u}}mmer, and
  Ulbrich}{Felsing et~al\mbox{.}}{2014}]%
        {felsign-verification-ase14}
\bibfield{author}{\bibinfo{person}{Dennis Felsing}, \bibinfo{person}{Sarah
  Grebing}, \bibinfo{person}{Vladimir Klebanov}, \bibinfo{person}{Philipp
  R{\"{u}}mmer}, {and} \bibinfo{person}{Mattias Ulbrich}.}
  \bibinfo{year}{2014}\natexlab{}.
\newblock \showarticletitle{{Automating regression verification}}. In
  \bibinfo{booktitle}{\emph{Proc. International Conference on Automated
  Software Engineering}}. \bibinfo{publisher}{ACM}, \bibinfo{pages}{349--360}.
\newblock
\urldef\tempurl%
\url{https://doi.org/10.1145/2642937.2642987}
\showDOI{\tempurl}


\bibitem[\protect\citeauthoryear{Godlin and Strichman}{Godlin and
  Strichman}{2009}]%
        {godlin-regression-verification-dac09}
\bibfield{author}{\bibinfo{person}{Benny Godlin} {and} \bibinfo{person}{Ofer
  Strichman}.} \bibinfo{year}{2009}\natexlab{}.
\newblock \showarticletitle{Regression verification}. In
  \bibinfo{booktitle}{\emph{Proc. Design Automation Conference}}.
  \bibinfo{publisher}{{ACM}}, \bibinfo{pages}{466--471}.
\newblock
\urldef\tempurl%
\url{https://doi.org/10.1145/1629911.1630034}
\showDOI{\tempurl}


\bibitem[\protect\citeauthoryear{Gulwani, Radicek, and Zuleger}{Gulwani
  et~al\mbox{.}}{2018}]%
        {gulwani-clustering-pldi18}
\bibfield{author}{\bibinfo{person}{Sumit Gulwani}, \bibinfo{person}{Ivan
  Radicek}, {and} \bibinfo{person}{Florian Zuleger}.}
  \bibinfo{year}{2018}\natexlab{}.
\newblock \showarticletitle{Automated clustering and program repair for
  introductory programming assignments}. In \bibinfo{booktitle}{\emph{Proc. ACM
  SIGPLAN Conference on Programming Language Design and Implementation}}.
  \bibinfo{publisher}{ACM}, \bibinfo{pages}{465--480}.
\newblock
\urldef\tempurl%
\url{https://doi.org/10.1145/3192366.3192387}
\showDOI{\tempurl}


\bibitem[\protect\citeauthoryear{Hur, Dreyer, Neis, and Vafeiadis}{Hur
  et~al\mbox{.}}{2012}]%
        {hur-bisimulation-popl12}
\bibfield{author}{\bibinfo{person}{Chung{-}Kil Hur}, \bibinfo{person}{Derek
  Dreyer}, \bibinfo{person}{Georg Neis}, {and} \bibinfo{person}{Viktor
  Vafeiadis}.} \bibinfo{year}{2012}\natexlab{}.
\newblock \showarticletitle{{The marriage of bisimulations and Kripke logical
  relations}}. In \bibinfo{booktitle}{\emph{Proc. Symposium on Principles of
  Programming Languages}}. \bibinfo{publisher}{ACM}, \bibinfo{pages}{59--72}.
\newblock
\urldef\tempurl%
\url{https://doi.org/10.1145/2103656.2103666}
\showDOI{\tempurl}


\bibitem[\protect\citeauthoryear{Jaber}{Jaber}{2020}]%
        {jaber-syteci-2020}
\bibfield{author}{\bibinfo{person}{Guilhem Jaber}.}
  \bibinfo{year}{2020}\natexlab{}.
\newblock \showarticletitle{SyTeCi: automating contextual equivalence for
  higher-order programs with references}.
\newblock \bibinfo{journal}{\emph{{PACMPL}}} \bibinfo{volume}{4},
  \bibinfo{number}{{POPL}} (\bibinfo{year}{2020}),
  \bibinfo{pages}{59:1--59:28}.
\newblock
\urldef\tempurl%
\url{https://doi.org/10.1145/3371127}
\showDOI{\tempurl}


\bibitem[\protect\citeauthoryear{Kaleeswaran, Santhiar, Kanade, and
  Gulwani}{Kaleeswaran et~al\mbox{.}}{2016}]%
        {kaleeswaran-coderassist-fse16}
\bibfield{author}{\bibinfo{person}{Shalini Kaleeswaran},
  \bibinfo{person}{Anirudh Santhiar}, \bibinfo{person}{Aditya Kanade}, {and}
  \bibinfo{person}{Sumit Gulwani}.} \bibinfo{year}{2016}\natexlab{}.
\newblock \showarticletitle{Semi-supervised verified feedback generation}. In
  \bibinfo{booktitle}{\emph{Proc. International Symposium on Foundations of
  Software Engineering}}. \bibinfo{publisher}{{ACM}},
  \bibinfo{pages}{739--750}.
\newblock
\urldef\tempurl%
\url{https://doi.org/10.1145/2950290.2950363}
\showDOI{\tempurl}


\bibitem[\protect\citeauthoryear{Koutavas and Wand}{Koutavas and Wand}{2006}]%
        {koutavas-bisimulation-popl06}
\bibfield{author}{\bibinfo{person}{Vasileios Koutavas} {and}
  \bibinfo{person}{Mitchell Wand}.} \bibinfo{year}{2006}\natexlab{}.
\newblock \showarticletitle{Small bisimulations for reasoning about
  higher-order imperative programs}. In \bibinfo{booktitle}{\emph{Proc.
  Symposium on Principles of Programming Languages}}. \bibinfo{publisher}{ACM},
  \bibinfo{pages}{141--152}.
\newblock
\urldef\tempurl%
\url{https://doi.org/10.1145/1111037.1111050}
\showDOI{\tempurl}


\bibitem[\protect\citeauthoryear{Liu, Wang, Wang, and Wu}{Liu
  et~al\mbox{.}}{2019}]%
        {liu-grading-formal-icse19}
\bibfield{author}{\bibinfo{person}{Xiao Liu}, \bibinfo{person}{Shuai Wang},
  \bibinfo{person}{Pei Wang}, {and} \bibinfo{person}{Dinghao Wu}.}
  \bibinfo{year}{2019}\natexlab{}.
\newblock \showarticletitle{Automatic grading of programming assignments: an
  approach based on formal semantics}. In \bibinfo{booktitle}{\emph{Proc.
  International Conference on Software Engineering: Software Engineering
  Education and Training}}. \bibinfo{publisher}{{IEEE} / {ACM}},
  \bibinfo{pages}{126--137}.
\newblock
\urldef\tempurl%
\url{https://doi.org/10.1109/ICSE-SEET.2019.00022}
\showDOI{\tempurl}


\bibitem[\protect\citeauthoryear{Perry, Kim, Samanta, and Zhang}{Perry
  et~al\mbox{.}}{2019}]%
        {perry-semcluster-pldi19}
\bibfield{author}{\bibinfo{person}{David~Mitchel Perry},
  \bibinfo{person}{Dohyeong Kim}, \bibinfo{person}{Roopsha Samanta}, {and}
  \bibinfo{person}{Xiangyu Zhang}.} \bibinfo{year}{2019}\natexlab{}.
\newblock \showarticletitle{{SemCluster: clustering of imperative programming
  assignments based on quantitative semantic features}}. In
  \bibinfo{booktitle}{\emph{Proc. ACM SIGPLAN Conference on Programming
  Language Design and Implementation}}. \bibinfo{pages}{860--873}.
\newblock
\urldef\tempurl%
\url{https://doi.org/10.1145/3314221.3314629}
\showDOI{\tempurl}


\bibitem[\protect\citeauthoryear{Pu, Narasimhan, Solar{-}Lezama, and
  Barzilay}{Pu et~al\mbox{.}}{2016}]%
        {pu-ml-splash16}
\bibfield{author}{\bibinfo{person}{Yewen Pu}, \bibinfo{person}{Karthik
  Narasimhan}, \bibinfo{person}{Armando Solar{-}Lezama}, {and}
  \bibinfo{person}{Regina Barzilay}.} \bibinfo{year}{2016}\natexlab{}.
\newblock \showarticletitle{sk{\_}p: a neural program corrector for MOOCs}. In
  \bibinfo{booktitle}{\emph{Proc. International Conference on Systems,
  Programming, Languages and Applications: Software for Humanity}}.
  \bibinfo{publisher}{ACM}, \bibinfo{pages}{39--40}.
\newblock
\urldef\tempurl%
\url{https://doi.org/10.1145/2984043.2989222}
\showDOI{\tempurl}


\bibitem[\protect\citeauthoryear{Singh, Gulwani, and Solar{-}Lezama}{Singh
  et~al\mbox{.}}{2013}]%
        {singh-autograder-pldi13}
\bibfield{author}{\bibinfo{person}{Rishabh Singh}, \bibinfo{person}{Sumit
  Gulwani}, {and} \bibinfo{person}{Armando Solar{-}Lezama}.}
  \bibinfo{year}{2013}\natexlab{}.
\newblock \showarticletitle{Automated feedback generation for introductory
  programming assignments}. In \bibinfo{booktitle}{\emph{Proc. ACM SIGPLAN
  Conference on Programming Language Design and Implementation}}.
  \bibinfo{publisher}{{ACM}}, \bibinfo{pages}{15--26}.
\newblock
\urldef\tempurl%
\url{https://doi.org/10.1145/2491956.2462195}
\showDOI{\tempurl}


\bibitem[\protect\citeauthoryear{Sumii and Pierce}{Sumii and Pierce}{2005}]%
        {pierce-bisimulation-popl05}
\bibfield{author}{\bibinfo{person}{Eijiro Sumii} {and}
  \bibinfo{person}{Benjamin~C. Pierce}.} \bibinfo{year}{2005}\natexlab{}.
\newblock \showarticletitle{A bisimulation for type abstraction and recursion}.
  In \bibinfo{booktitle}{\emph{Proc. Symposium on Principles of Programming
  Languages}}. \bibinfo{publisher}{ACM}, \bibinfo{pages}{63--74}.
\newblock
\urldef\tempurl%
\url{https://doi.org/10.1145/1040305.1040311}
\showDOI{\tempurl}


\bibitem[\protect\citeauthoryear{Wang, Singh, and Su}{Wang
  et~al\mbox{.}}{2018}]%
        {wang-sarfgen-pldi18}
\bibfield{author}{\bibinfo{person}{Ke Wang}, \bibinfo{person}{Rishabh Singh},
  {and} \bibinfo{person}{Zhendong Su}.} \bibinfo{year}{2018}\natexlab{}.
\newblock \showarticletitle{Search, align, and repair: data-driven feedback
  generation for introductory programming exercises}. In
  \bibinfo{booktitle}{\emph{Proc. ACM SIGPLAN Conference on Programming
  Language Design and Implementation}}. \bibinfo{publisher}{ACM},
  \bibinfo{pages}{481--495}.
\newblock
\urldef\tempurl%
\url{https://doi.org/10.1145/3192366.3192384}
\showDOI{\tempurl}


\bibitem[\protect\citeauthoryear{Zuck, Pnueli, Fang, and Goldberg}{Zuck
  et~al\mbox{.}}{2002}]%
        {zpf+translation-2002}
\bibfield{author}{\bibinfo{person}{Lenore Zuck}, \bibinfo{person}{Amir Pnueli},
  \bibinfo{person}{Yi Fang}, {and} \bibinfo{person}{Benjamin Goldberg}.}
  \bibinfo{year}{2002}\natexlab{}.
\newblock \showarticletitle{VOC: A translation validator for optimizing
  compilers}.
\newblock \bibinfo{journal}{\emph{Electronic notes in theoretical computer
  science}} \bibinfo{volume}{65}, \bibinfo{number}{2} (\bibinfo{year}{2002}),
  \bibinfo{pages}{2--18}.
\newblock
\urldef\tempurl%
\url{https://doi.org/10.1016/S1571-0661(04)80393-1}
\showDOI{\tempurl}


\end{thebibliography}

\end{document}